\documentclass[a4paper,11pt,twoside]{article} 
\pdfoutput=1
\usepackage{graphicx}
\usepackage{epstopdf}
\DeclareGraphicsExtensions{.jpg,.pdf,.png}
\graphicspath{{../FIG/}}
\begin{document}

\title{The Cosmic Microwave Background for Pedestrians: \\ A Review for Particle and Nuclear Physicists}
\author{Dorothea Samtleben$^{1)}$
Suzanne Staggs $^{2)}$
Bruce Winstein $^{3)}$
}
\date{\small 1) Max-Planck-Institut f\"ur Radioastronomie, Bonn~~2) Princeton University \\3) The University of Chicago}
\markboth{{Samtleben,} {Staggs \&} {Winstein}}{The CMB for Pedestrians}
\thispagestyle{empty}
\maketitle
\setcounter{tocdepth}{2}
\tableofcontents
\begin{abstract}
We intend to show how fundamental science is drawn from the patterns
in the temperature and polarization fields of the cosmic microwave
background (CMB) radiation, and thus to motivate the field of CMB
research. We discuss the field's history, potential science and
current status, contaminating foregrounds, detection and analysis
techniques and future prospects. Throughout the review we draw
comparisons to particle physics, a field that has many of the same goals
and that has gone through many of the same stages.
\end{abstract}
\maketitle

\pagenumbering{roman}  
\newcounter{save_roman}
\setcounter{save_roman}{\value{page}}
\pagenumbering{arabic}
\setcounter{page}{1}    

\section{INTRODUCTION}
What is all the fuss about noise? In this review we endeavor to convey
the excitement and promise of studies of the cosmic microwave
background (CMB) radiation to scientists not engaged in these studies,
particularly to particle and nuclear physicists. Although the
techniques for both detection and data processing are quite far apart
from those familiar to our intended audience, the science goals are
aligned. We do not emphasize mathematical rigor \footnote{This review
complements the one by Kamionkowski and Kosowsky (1).}, but rather
attempt to provide insight into (a) the processes that allow
extraction of fundamental physics from the observed radiation patterns
and (b) some of the most fruitful methods of detection.  

In Section 1, we begin with a broad outline of the most relevant
physics that can be addressed with the CMB and its polarization. We
then treat the early history of the field, how the CMB and its
polarization are described, the physics behind the acoustic peaks, and
the cosmological physics that comes from CMB studies. Section~2
presents the important foreground problem: primarily galactic sources
of microwave radiation. The third section treats detection techniques
used to study these extremely faint signals. The promise (and
challenges) of future studies is presented in the last two
sections. In keeping with our purposes, we do not cite an exhaustive
list of the ever expanding literature on the subject, but rather
indicate several particularly pedagogical works.
\subsection{The Standard Paradigm}
Here we briely review the now standard framework in which cosmologists
work and for which there is abundant evidence. We recommend readers to
the excellent book Modern Cosmology by Dodelson (2). Early in its
history (picoseconds after the Big Bang), the energy density of the
Universe was divided among matter, radiation, and dark energy. The
matter sector consisted of all known elementary particles and included
a dominant component of dark matter, stable particles with negligible
electromagnetic interactions. Photons and neutrinos (together with the
kinetic energies of particles) comprised the radiation energy density,
and the dark energy component some sort of fluid with a negative
pressure appears to have had no importance in the early Universe,
although it is responsible for its acceleration today. 

Matter and radiation were in thermal equilibrium, and their combined
energy density drove the expansion of space, as described by general
relativity. As the Universe expanded, wavelengths were stretched so
that particle energies (and hence the temperature of the Universe)
decreased: T($z$) = T(0)(1 + $z$), where $z$ is the redshift and T(0)
is the temperature at $z$ = 0, or today. There were slight
overdensities in the initial conditions that, throughout the
expansion, grew through gravitational instability, eventually forming
the structure we observe in todays Universe: myriad stars, galaxies,
and clusters of galaxies. 

The Universe was initially radiation dominated. Most of its energy
density was in photons, neutrinos, and kinetic motion. After the
Universe cooled to the point at which the energy in rest mass equaled
that in kinetic motion (matter-radiation equality), the expansion rate
slowed and the Universe became matter dominated, with most of its
energy tied up in the masses of slowly moving, relatively heavy stable
particles: the proton and deuteron from the baryon sector and the dark
matter particle(s). The next important era is termed either
decoupling, recombination, or last scattering. When the temperature
reached roughly 1 eV, atoms (mostly H) formed and the radiation cooled
too much to ionize. The Universe became transparent, and it was during
this era that the CMB we see today was emitted, when physical
separations were 1000 times smaller than today ($z\approx$1000). At
this point, less than one million years into the expansion, when
electromagnetic radiation ceased playing an important dynamical role,
baryonic matter began to collapse and cool, eventually forming the
first stars and galaxies. Later, the first generation of stars and
possibly supernova explosions seem to have provided enough radiation
to completely reionize the Universe (at $z\approx 10$). Throughout
these stages, the expansion was decelerated by the gravitational force
on the expanding matter. Now we are in an era of cosmic acceleration
($z\leq$2), where we find that approximately 70\% of the energy
density is in the fluid that causes the acceleration, 25\% is in dark
matter, and just 5\% is in baryons, with a negligible amount in
radiation.
\subsection{Fundamental Physics in the CMB}
The CMB is a record of the state of the Universe at a fraction of a
million years after the Big Bang, after a quite turbulent beginning,
so it is not immediately obvious that any important information
survives. Certainly the fundamental information available in the
collisions of elementary particles is best unraveled by observations
within nanoseconds of the collision. Yet even in this remnant
radiation lies the imprint of fundamental features of the Universe at
its earliest moments. 
\subsubsection{CMB features: evidence for inflation.} 
One of the most important features of the CMB is its Planck
spectrum. It follows the blackbody curve to extremely high precision,
over a factor of approximately 1000 in frequency (see Figure~1). This
implies that the Universe was in thermal equilibrium when the
radiation was released (actually long before, as we see below), which
was at a temperature of approximately 3000 K. Today it is near 3 K.

An even more important feature is that, to better than a part in
10$^4$, this temperature is the same over the entire sky. This is
surprising because it strongly implies that everything in the
observable Universe was in thermal equilibrium at one time in its
evolution. Yet at any time and place in the expansion history of the
Universe, there is a causal horizon defined by the distance light (or
gravity) has traveled since the Big Bang; at the decoupling era, this
horizon corresponded to an angular scale of approximately 1$^\circ$,
as observed today. The uniformity of the CMB on scales well above
1$^\circ$ is termed the horizon problem. 

The most important feature is that there are differences in the CMB
temperature from place to place, at the level of 10$^{-5}$, and that
these fluctuations have coherence beyond the horizon at the time of
last scattering. The most viable notion put forth to address these
observations is the inflationary paradigm, which postulates a very
early period of extremely rapid expansion of the Universe. Its scale
factor increased by approximately 21 orders of magnitude in only
approximately 10$^{35}$ s. Before inflation, the small patch that
evolves into our observable Universe was likely no larger across than
the Planck length, its contents in causal contact and local
thermodynamic equilibrium. The process of superluminal inflation
disconnects regions formally in causal contact. When the expansion
slowed, these regions came back into the horizon and their initial
coherence became manifest. 

The expansion turns quantum fluctuations into (nearly) scale-invariant
CMB inhomogeneities, meaning that the fluctuation power is nearly the
same for all threedimensional Fourier modes. So far, observations
agree with the paradigm, and scientists in the field use it to
organize all the measurements.  Nevertheless, we are far from
understanding the microphysics driving inflation. The number of models
and their associated parameter spaces greatly exceed the number of
relevant observables. New observations, particularly of the CMB
polarization, promise a more direct look at inflationary physics,
moving our understanding from essentially kinematical to dynamical.
\subsubsection{Probing the Universe when T = 10$^{16}$~GeV.} 
For particle physicists, probing microphysics at energy scales beyond
accelerators using cosmological observations is attractive. The
physics of inflation may be associated with the grand unifcation
scale, and if so, there could be an observable signature in the CMB:
gravity waves. Metric perturbations, or gravity waves (also termed
tensor modes), would have been created during inflation, in addition
to the density perturbations (scalar modes) that give rise to the
structure in the Universe today. 

In the simplest of inflationary models, there is a direct relation
between the energy scale of inflation and the strength of these
gravity waves. The notion is that the Universe initially had all its
energy in a scalar field $\Phi$ displaced from the minimum of its
potential $V$. $V(\Phi)$ is suitably constructed so that slowly rolls
down its potential, beginning the inflationary era of the Universe,
which terminates only when $\Phi$ approaches its minimum. 

Inflation does not predict the level of the tensor (or even scalar)
modes. The parameter $r = T/S$ is the tensor-to-scalar ratio for
fluctuation power; it depends on the energy scale at which inflation
began. Specifically, the initial height of the potential $V_i$ depends
on $r$, as $V_i = r(0.003 M_{pl})^4$. A value of $r = 0.001$, perhaps
the smallest detectable level, corresponds to $V_i^{0.25} = 6.5 \times
10^{15} GeV$. 

The tensor modes leave distinct patterns on the polarization of the
CMB, which may be detectable. This is now the most important target
for future experiments. They also have effects on the temperature
anisotropies, which currently limit $r$ to less than approximately 0.3.
\subsubsection{How neutrino masses affect the CMB.}
It is a remarkable fact that even a slight neutrino mass affects
the expansion of the Universe. When the dominant dark matter clusters,
it provides the environment for baryonic matter to collapse, cool, and
form galaxies. As described above, the growth of these structures
becomes more rapid in the matter-dominated era. If a significant
fraction of the dark matter were in the form of neutrinos with
electron-volt-scale masses (nonrelativistic today), these would have
been relativistic late enough in the expansion history that they could
have moved away from overdense regions and suppressed structure
growth. Such suppression alters the CMB patterns and provides some
sensitivity to the sum of the neutrino masses. Note also that
gravitational effects on the CMB in its passage from the epoch (or
surface) of last scattering to the present leave signatures of that
structure and give an additional (and potentially more sensitive)
handle on the neutrino masses (see Section 1.9.2). 
\subsubsection{Dark energy.} 
We know from the CMB that the geometry of the Universe is consistent
with being flat. That is, its density is consistent with the critical
density. However, the overall density of matter and radiation
discerned today (the latter from the CMB directly) falls short of
accounting for the critical density by approximately a factor of
three, with little uncertainty. Thus, the CMB provides indirect
evidence for dark energy, corroborating supernova studies that
indicate a new era of acceleration. Because the presence and possible
evolution of a dark energy component alter the expansion history of
the Universe, there is the promise of learning more about this
mysterious component. 
\subsection{History}
In 1965, Penzias and Wilson (3), in trying to understand a nasty noise
source in their experiment to study galactic radio emission,
discovered the CMB arguably the most important discovery in all the
physical sciences in the twentieth century. Shortly thereafter,
scientists showed that the radiation was not from radio galaxies or
reemission of starlight as thermal radiation. This first measurement was
made at a central wavelength of 7.35 cm, far from the blackbody
peak. The reported temperature was $T = 3.5 \pm 1 K$. However, for a
blackbody, the absolute flux at any known frequency determines its
temperature. Figure~1 shows the spectrum of detected radiation for
different temperatures. There is a linear increase in the peak
position and in the flux at low frequencies (the Rayleigh-Jeans part
of the spectrum) as temperature increases. 
\begin{figure}[h]
\begin{center}
\includegraphics[scale=.50,angle=180]{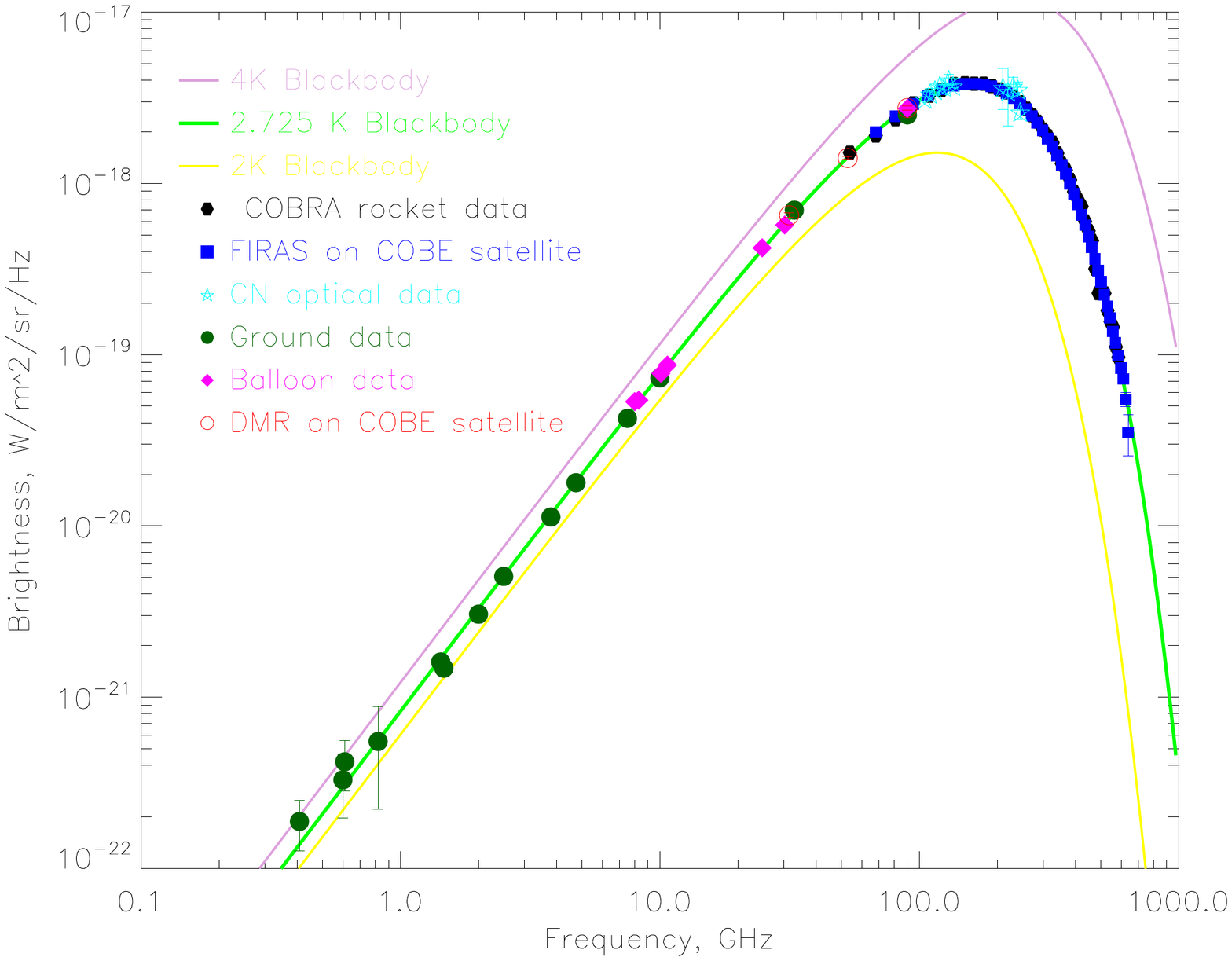}
\caption[The Planck Spectrum of the CMB]{Measurements of the CMB flux
vs. frequency together with a fit to the data.  Superposed are the
expected black body curves for T = 2~K and T = 40K.\label{Planck} }
\end{center}
\end{figure}

Multiple efforts were soon mounted to confirm the blackbody nature of
the CMB and to search for its anisotropies. Partridge (4) gives a very
valuable account of the early history of the field. However, there
were false observations, which was not surprising given the low ratio
of signal to noise. Measurements of the absolute CMB temperature are
at milli-Kelvin levels, whereas relative measurements between two
places on the sky are at micro-Kelvin levels. By 1967, Partridge and
Wilkinson had shown, over large regions of the sky, that $\Delta T/T
\leq (1 - 3) \times 10^{-3}$, leading to the conclusion that the
Universe was in thermal equilibrium at the time of decoupling
(4). However, nonthermal injections of energy even at much earlier
times, for example, from the decays of long-lived relic particles,
would distort the spectrum. It is remarkable that current precise
measurements of the blackbody spectrum can push back the time of
significant injections of energy to when the Universe was barely a
month old (5). Thus, recent models that attribute the dark matter to
gravitinos as decay products of long-lived supersymmetric weakly
interacting massive particles (SUSY WIMPs) (6) can only tolerate
lifetimes of less than approximately one month. 

The solar system moves with velocity $\beta \approx 3 \times 10^{-3}$,
causing a dipole anisotropy of a few milli-Kelvins, first detected in
the 1980s. (Note that the direction of our motion was not the one
initially hypothesized from motions of our local group of galaxies.)
The first detection of primordial anisotropy came from the COBE
satellite (7) in 1992, at the level of 10$^{-5}$ (30 $\mu$K), on
scales of approximately 10$^\circ$ and larger. The impact of this
detection matched that of the initial discovery. It supported the idea
that structure in the Universe came from gravitational instability to
overdensities. The observed anisotropies are a combination of the
original ones at the time of decoupling and the subsequent
gravitational red- or blueshifting as photons leave over- or
underdense regions.
\subsection{Introduction to the Angular Power Spectrum}
Here we describe the usual techniques for characterizing the
temperature field. First, we define the normalized temperature $\Theta$ in
direction $\hat{\bf n}$ on the celestial sphere by the deviation $\Delta T$ from the
average: $\Theta(\hat{\bf n}) = \frac{\Delta T}{<T>}$. Next, we consider
the multipole decomposition of this temperature field in terms of
spherical harmonics $Y_{lm}$:
\begin{equation} \Theta_{lm} = \int \Theta(\hat{\bf n})Y_{lm}^*(\hat{\bf n}) d\Omega
,\end{equation} where the integral is over the entire sphere.

If the sky temperature field arises from Gaussian random fluctuations,
then the field is fully characterized by its power spectrum
$\Theta^*_{lm}\Theta_{l'm'}$. The order $m$ describes the angular
orientation of a fluctuation mode, but the degree (or multipole) $l$
describes its characteristic angular size. Thus, in a Universe with no
preferred direction, we expect the power spectrum to be independent of
$m$. Finally, we define the angular power spectrum $C_l$ by
$<\Theta_{lm}^*\Theta_{l'm'}>= \delta_{ll'}\delta_{mm'} C_l$ . Here
the brackets denote an ensemble average over skies with the same
cosmology. The best estimate of $C_l$ is then from the average over
m. 

Because there are only the ($2l + 1$) modes with which to detect the
power at multipole $l$, there is a fundamental limit in determining
the power.  This is known as the cosmic variance (just the variance on
the variance from a finite number of samples):
\begin{equation}\frac{\Delta C_l}{Cl} = \sqrt{\frac{2}{2l + 1}}.\end{equation}
The full uncertainty in the power in a given multipole degrades from
instrumental noise, finite beam resolution, and observing over a finite
fraction of the full sky, as shown below in Equation 9. 

For historical reasons, the quantity that is usually plotted,
sometimes termed the TT (temperature-temperature correlation)
spectrum, is
\begin{equation}\Delta T^2 \equiv \frac{l(l + 1)}{2\pi} C_l T_{CMB}^2,\end{equation}
where $T_{CMB}$ is the blackbody temperature of the CMB. This
is the variance (or power) per logarithmic interval in $l$ and is
expected to be (nearly) uniform in inflationary models (scale
invariant) over much of the spectrum. This normalization is useful in
calculating the contributions to the fluctuations in the temperature
in a given pixel from a range of $l$ values: 
\begin{equation}\Delta T^2 = \int^{l_{max}}_{l_{min}}\frac{(2l + 1)}{4\pi} C_l T_{CMB}^2 dl\end{equation}
\begin{figure}[tbp]
\begin{center}
\includegraphics[scale=.3]{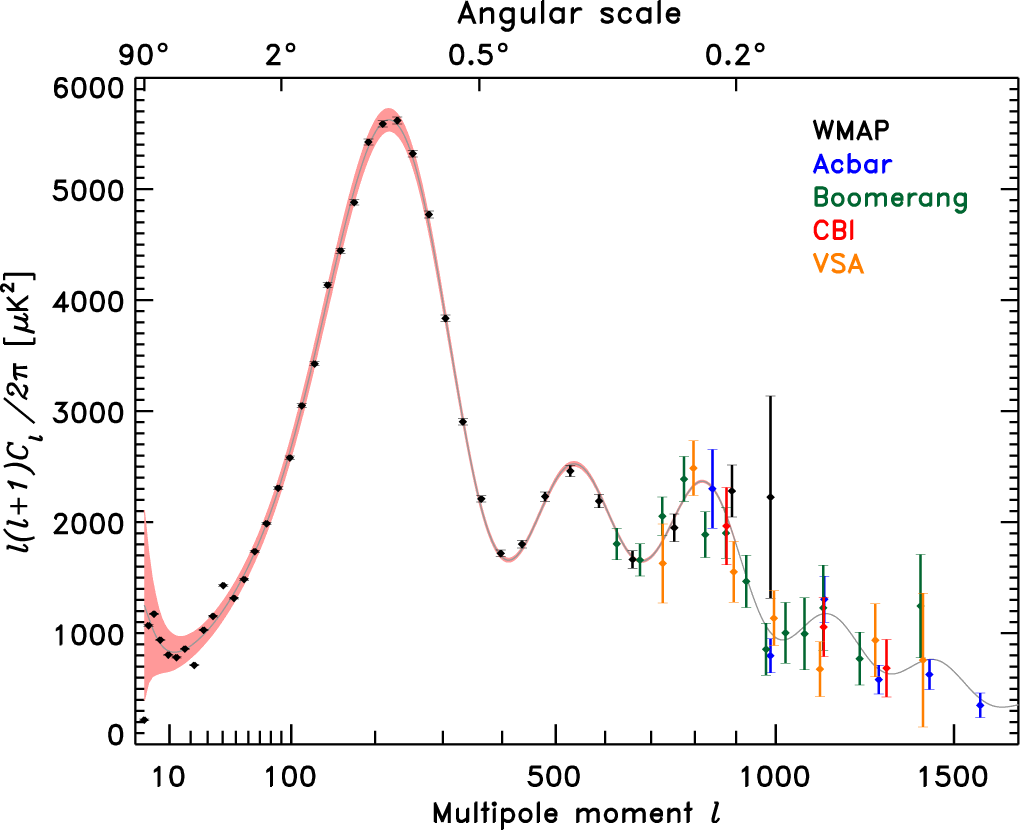}
 \caption[The TT Power Spectrum] {The TT Power Spectrum.  Data from
the Wilkinson Microwave Anisotropy Probe (WMAP) (8), and high-$l$ data
from other experiments are shown, in addition to the best-fit cosmological
model to the WMAP data alone. Note the multipole scale on the bottom
and the angular scale on the top. Figure courtesy of the WMAP science team\label{WMAP-PS}}
\end{center}
\end{figure}
\subsection{Current Understanding of the Temperature Field} 
Figure 2 shows the current understanding of the temperature power
spectrum (from herewith we redefine $C_l$ to have $K^2$ units by
replacing $C_l$ with $C_l T_{CMB}^2$). The region below $l \approx 20$
indicates the initial conditions. These modes correspond to Fourier
modes at the time of decoupling, with wavelengths longer than the
horizon scale. Note that were the sky describable by random white
noise, the $C_l$ spectrum would be flat and the TT power spectrum,
defined by Equation 3, would have risen in this region like $l^2$. The
(pleasant) surprise was the observation of finite power at these
superhorizon scales. At high $l$ values, there are acoustic
oscillations, which are damped at even higher $l$ values. The
positions and heights of the acoustic-oscillation peaks reveal
fundamental properties about the geometry and composition of the
Universe, as we discuss below.
\subsection{Acoustic Oscillations}
The CMB data reveal that the initial inhomogeneities in the Universe
were small, with overdensities and underdensities in the dark matter,
protons, electrons, neutrinos, and photons, each having the
distribution that would arise from a small adiabatic compression or
expansion of their admixture. An overdense region grows by attracting
more mass, but only after the entire region is in causal contact. 

We noted that the horizon at decoupling corresponds today to
approximately 1$^\circ$ on the sky. Only regions smaller than this had
time to compress before decoupling. For sufficiently small regions,
enough time elapses that compression continues until the photon
pressure is sufficient to halt the the electrons via Thomson
scattering, and the protons follow the electrons to keep a charge
balance. Inflation provides the initial conditions - zero velocity.

Decoupling preserves a snapshot of the state of the photon fluid at
that time. Excellent pedagogical descriptions of the oscillations can
be found at {\it http://background.uchicago.edu/$\sim$whu/}. Other
useful pages are {\it
http://wmap.gsfc.nasa.gov/,http://space.mit.edu/home/tegmark/index.html}
and {\it http://www.astro.ucla.edu/$\sim$wright/intro.html}.
Perturbations of particular sizes may have undergone (a) one
compression, (b) one compression and one rarefaction, (c) one
compression, one rarefaction, and one compression, and so on. Extrema
in the density field result in maxima in the power spectrum. 

Consider a standing wave permeating space with frequency $\omega$ and
wave number $k$, where these are related by the velocity of
displacements (the sound speed, $v_s \approx c/\sqrt{3}$) in the
plasma: $\omega = k v_s$ . The wave displacement $A_k$ for this single
mode can then be written as $A_k(x,t) \propto sin(kx) cos(\omega
t)$. The displacement is maximal at time $t_{dec}$ of decoupling for
$k_{TT} v_s t_{dec} = \pi, 2\pi, 3\pi ...$ We add the TT subscript to
label these wave numbers associated with maximal autocorrelation in
the temperature. Note that even in this tightly coupled regime, the
Universe at decoupling was quite dilute, with a physical density of
less than $10^{-20} g~cm^{-3}$. Because the photons diffuse their mean
free path is not infinitely short this pattern does not go on without
bound. The overtones are damped, and in practice only five or six such
peaks will be observed, as seen in Figure~2.
\subsection{How Spatial Modes Look Like Angular Anisotropies} 
To help explain these ideas, we reproduce a few frames from an
animation by W. Hu. Figure 3 shows a density fluctuation on the sky
from a single $k$ mode and how it appears to an observer at different
times. The figure shows the particle horizon just after decoupling. This
represents the farthest distance one could in principle see
approximately the speed of light times the age of the Universe. An
observer at the center of the gure could not by any means have
knowledge of anything outside this region. Of course, just after
decoupling, the observer could see a far shorter distance. Only then
could light propagate freely. 

The subsequent frames show how the particle horizon grows to encompass
more corrugations of the original density fluctuation. At first the
observer sees a dipole, later a quadrupole, then an octopole, and so
on, until the present time when that single mode in density
inhomogeneities creates very high multipoles in the temperature
anisotropy. 

It is instructive to think of how the temperature observed today at a
spot on the sky arises from the local moments in the temperature field
at the time of last scattering. It is only the lowest three moments
that contribute to determining the anisotropies. The monopole terms
are the ones transformed into the rich angular spectra. The dipole
terms also have their contribution: The motion in the fluid
oscillations results in Doppler shifts in the observed
temperatures. Polarization, we see below, comes from local
quadrupoles.
\begin{figure}[htbp]
\includegraphics[scale=.22]{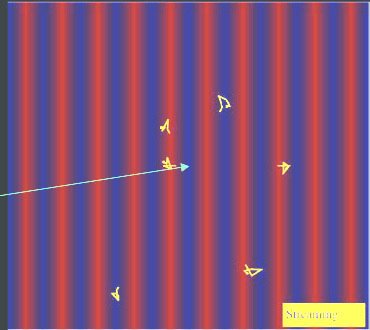}
\includegraphics[scale=.22]{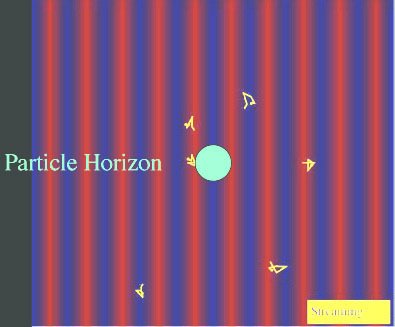}
\includegraphics[scale=.22]{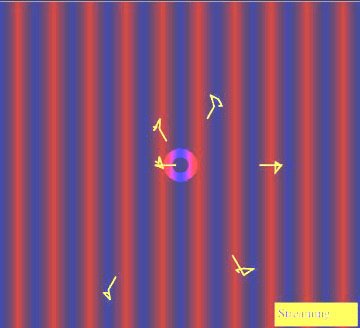}
\includegraphics[scale=.22]{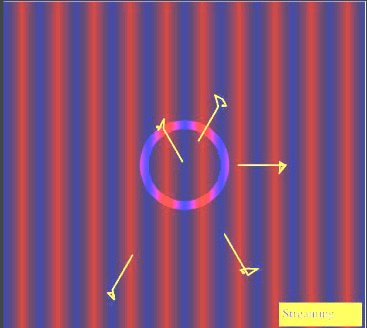}
\includegraphics[scale=.22]{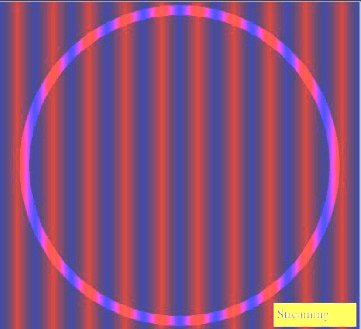}
\caption{The Signature of one frozen mode after Decoupling.}{These
frames show one superhorizon temperature mode just after decoupling
with representative photons last scattering and heading toward the
observer at the center. Left to right: just after decoupling; the
observer's particle horizon when only the temperature monopole can be
detected; som e time later when the quadrupole is detected; later
still when the 12-pole is detected; and today, a very high, well
aligned multipole, from just this single mode in $k$-space, is
detected. Figure courtesy of W. Hu \label{Hu-1}}
\end{figure}
\subsubsection{Inflation revisited.} Inflation is a mechanism
whereby fluctuations are created without violating causality. There
does not seem to be a better explanation for the observed
regularities. Nevertheless, Wolfgang Pauli's famous statement about
the neutrino comes to mind: I have done a terrible thing: I have
postulated a particle that cannot be detected!  

Sometimes it seems that inflation is an idea that cannot be tested, or
tested incisively. Of course Pauli's neutrino hypothesis did test
positive, and similarly there is hope that the idea of inflation can
reach the same footing. Still, we have not (yet) seen any scalar field
in nature. We discuss what has been claimed as the smoking gun test of
inflation the eventual detection of gravity waves in the CMB. However,
will we ever know with certainty that the Universe grew in volume by a
factor of 10$^{63}$ in something like $10^{-35} s$?
\subsection{CMB Polarization}
Experiments have now shown that the CMB is polarized, as
expected. Researchers now think that the most fruitful avenue to
fundamental physics from the CMB will be in precise studies of the
patterns of the polarization. This section treats the mechanisms
responsible for the generation of the polarization and how this
polarization is described. 
\subsubsection{How polarization gets generated}
If there is a quadrupole anisotropy in the temperature field around a
scattering center, even if that radiation is unpolarized, the
scattered radiation will be as shown in Figure 4: A linear
polarization will be generated.
\begin{figure} 
\includegraphics[width=1.7in]{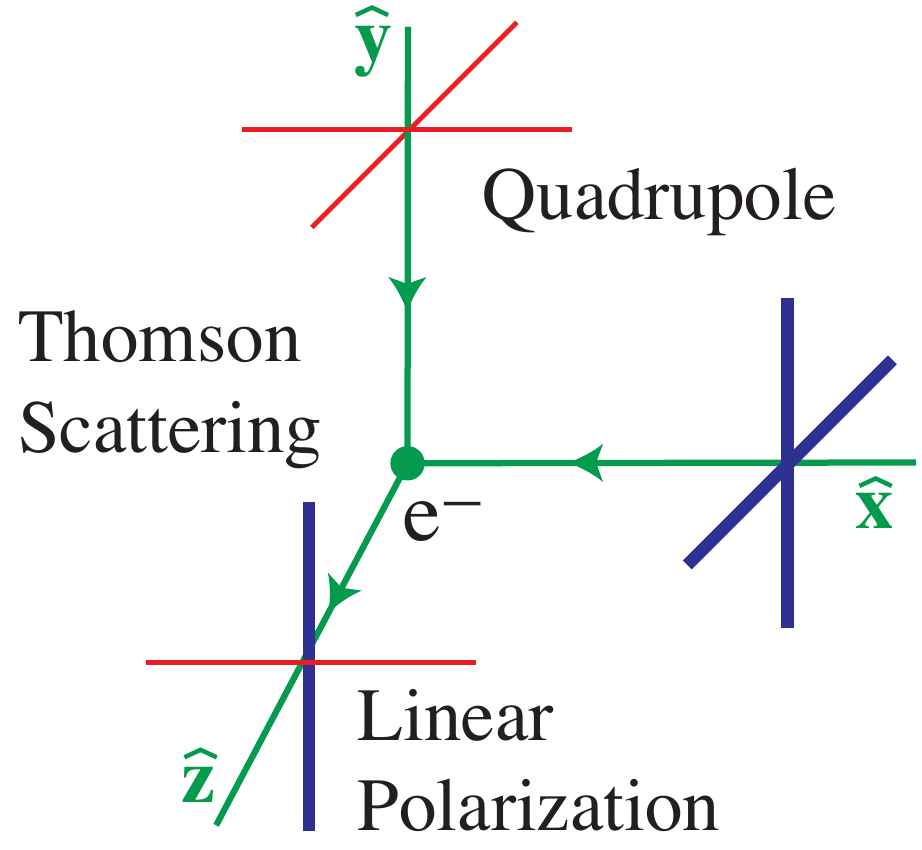}
\hspace*{1cm}\includegraphics[scale=0.7]{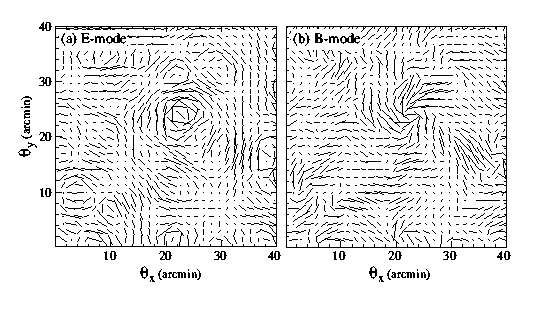}
\caption{Generation of polarization. Left: Unpolarized but anisotropic
radiation incident on an electron produces polarized
radiation. Intensity is represented by line thickness. To an observer
looking along the direction of the scattered photons ($\hat{z}$), the
incoming quadrupole pattern produces linear polarization along the
$\hat{y}$-direction. In terms of the Stokes parameters, this is
$Q=(E_x^2 - E_y^2)/2$, the power difference detected along the
$\hat{x}$- and $\hat{y}$-directions. Linear polarization needs one
other parameter, corresponding to the power difference between
45$^\circ$ and 135$^\circ$ from the x-axis. This parameter is easily
shown to be Stokes $U=E_x E_y$. Right: E and B polarization patterns. The
length of the lines represents the degree of polarization, while their
orientation gives the direction of maximum electric field. Frames
courtesy of W. Hu.}
\end{figure}
The quadrupole is generated during decoupling, as shown in Figure
3. Because the polarization arises from scattering but said scattering
dilutes the quadrupole, the polarization anisotropy is much weaker
than that in the temperature field. Indeed with each scatter on the
way to equilibrium, the polarization is reduced. Any remaining
polarization is a direct result of the cessation of scattering. For
this reason, the polarization peaks at higher $l$ values than does the
temperature anisotropy. The local quadrupole on scales that are large
in comparison to the mean free path is diluted from multiple
scattering. 
\subsubsection{The E and B polarization fields.}
The polarization field is both more complicated and richer than the
temperature field. At each point in the sky, one must specify both the
degree of polarization and the preferred direction of the electric
field. This is a tensor field that can be decomposed into two types,
termed E and B, which are, respectively, scalar and pseudoscalar fields,
with associated power spectra. Examples of these polarization fields are
depicted schematically in Figure~4. The E and B fields are more
fundamental than the polarization field on the sky, whose description is
coordinate-system dependent. In addition, E modes arise from the
density perturbations (which do not produce B modes) that we describe,
whereas the B modes come from the tensor distortions to the space-time
metric (which do have a handedness). We mention here that the E and B
fields are nonlocal. Their extraction from measurements of polarization
over a set of pixels, often in a finite patch of sky, is a
well-developed but subtle procedure (see Section 3.3). 

The peaks in the EE (E-polarization correlated with itself ) spectrum
should be 180æ out of phase with those for temperature: Polarization
results from scattering and thus is maximal when the fluid velocity is
maximal. Calculating the fluid velocity for the mode in Section 1.6,
we find $k_{EE} v_s t_{dec} = \pi/2, 3\pi/2, 5\pi/2 ...$ , defining
modes with maximal EE power. The TE (E-polarization correlated with
the temperature field) spectrum how modes in temperature correlate
with those with E polarization is also of cosmological interest, with
its own peak structure. Here we are looking at modes that have a
maximum at decoupling in the product of their temperature and E-mode
polarization (or velocity). Similarly, the appropriate maxima (which
in this case can be positive or negative) are obtained when $k_{TE}
v_s t_{dec} = \pi/4, 3\pi/4, 5\pi/4 ...$ Thus, between every peak in
the TT power spectrum there should be one in the EE, and between every
TT and EE pair of peaks there should be one in the TE.
\subsubsection{Current understanding of polarization data}
Figure 5 shows the EE results in addition to the expected power
spectra in the standard cosmological model. Measurements of the TE
cross correlation are also shown. The pattern of peaks in both power
spectra is consistent with what was expected. What was unexpected was
the enhancement at the lowest $l$ values in the EE power spectrum. This
is discussed in the next section. 

The experiments reported in Figure~5, with 20 or fewer detectors, use
a variety of techniques and operate in different frequency
ranges. This is important in dealing with astrophysical foregrounds
(see Section 2) that have a different frequency dependence from that
of the CMB.
\begin{figure}[h]
\hspace*{-3cm}
\includegraphics[scale=0.35, angle=90]{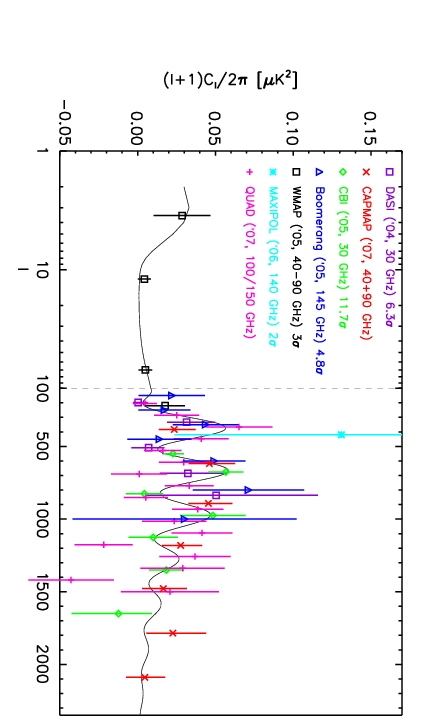}
\includegraphics[scale=0.35, angle=90]{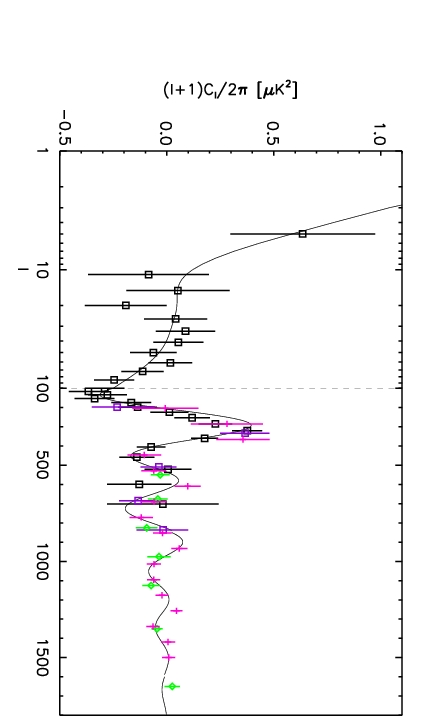}
\caption[Polarization Results.]{Measurements of EE (left) and TE
(right) power spectra together with the WMAP best-fit cosmological
model. The names of the experiments, their years of publication, and
the frequency ranges covered are indicated, as well as the number of
standard deviations with which each experiment claims a
detection. Note the change from logarithmic to linear multipole scale
at $l = 100$ and that to display features in the very low $l$ range, we
plot $(l + 1)C_l/2\pi$.}
\end{figure} 
Limits from current experiments on the B-mode power are now at the
level of 1-10 $\mu K^2$, far from the expected signal levels shown in
Figure 6. The peak in the power spectrum (for the gravity waves) is at
$l \approx 100$, the horizon scale at decoupling. The reader may wonder
why the B modes fall off steeply above this scale and show no acoustic
oscillations. The reason is simple: A tensor mode will give, for
example, a compression in the x-direction followed by a rarefaction in
the y-direction, but will not produce a net overdensity that would
subsequently contract. In the final section we discuss experiments with
far greater numbers of detectors aimed specifically at B-mode
science. Note that such gravity waves have frequencies today of order
$10^{-16}$~Hz. However, if their spectrum approximates one of scale
invariance, they would in principle be detectable at frequencies
nearer 1 Hz, such as in the LISA experiment. This is discussed more
fully in Reference 10.  
\subsection{Processes after Decoupling: Secondary Anisotropies}
In this section we briefly discuss three important processes after
decoupling: rescattering of the CMB in the reionized plasma of the
Universe, lensing of the CMB through gravitational interactions with
matter, and scattering of the CMB from hot gas in Galaxy
clusters. Although these can be considered foregrounds perturbing the
primordial information, each can potentially provide fundamental
information. 
\begin{figure}[h]
\vspace*{-6cm}
\includegraphics[scale=0.5]{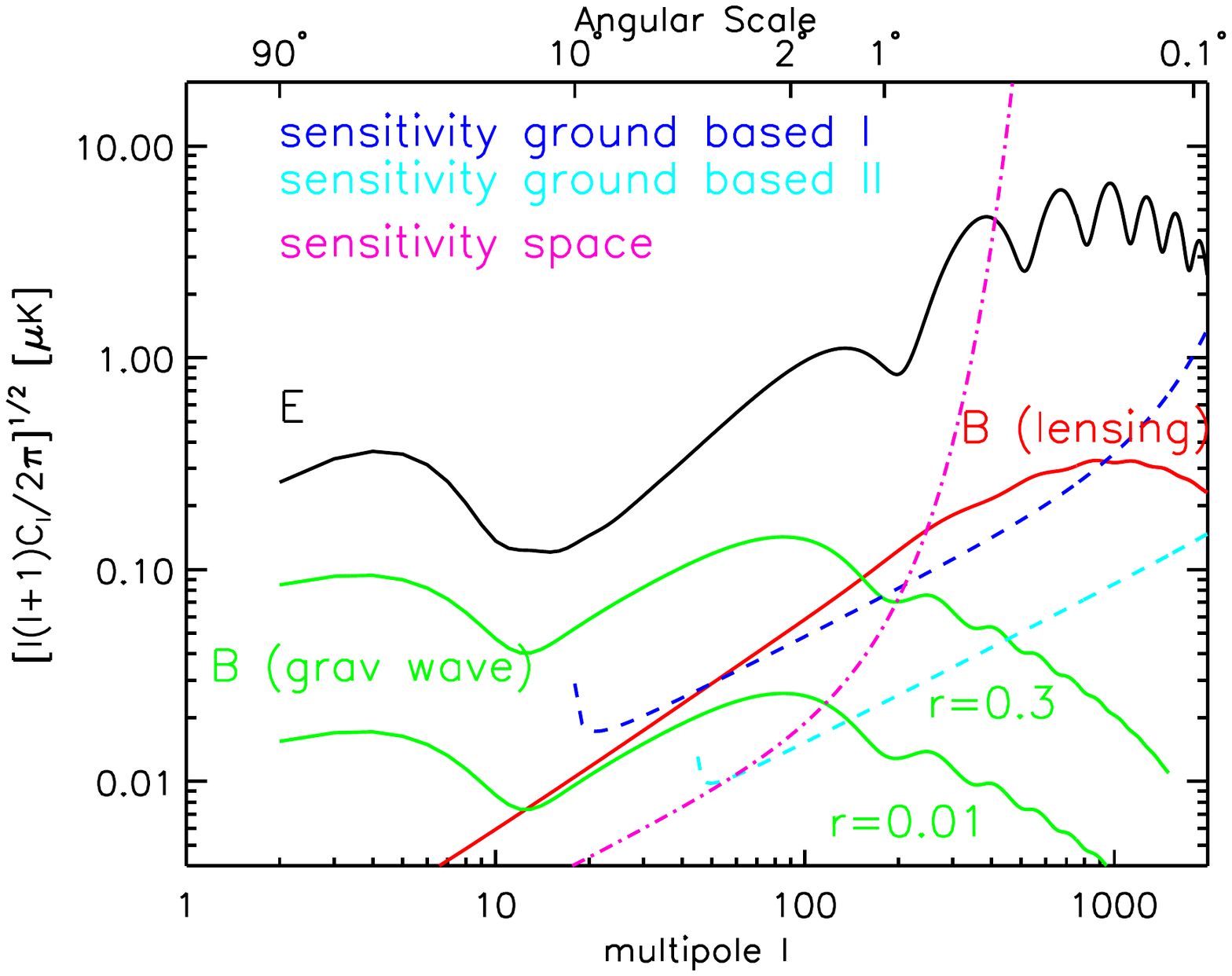}
\caption{CMB polarization power spectra and estimated sensitivity of
future experiments. The solid curves show the predictions for the E-
and B-mode power spectra. The primordial B-mode power spectrum is
shown for $r = 0.3$ and $r = 0.01$. The predicted B-mode signal power
spectrum due to the distortion of E modes by weak gravitational
lensing is also shown. Estimated statistical sensitivities for a new
space mission (pink line) and two sample ground-based experiments, as
considered in Reference 9, each with 1000 detectors operating for one
year with 100\% duty cycle (dark and light blue lines), are shown.
Experiment I observes 4\% of the sky, with a 6-arcmin resolution;
experiment II observes 0.4\% of the sky, with a 1-arcmin
resolution.}
\end{figure}
\subsubsection{Reionization.} 
The enhancement in the EE power spectrum at the very lowest $l$ values
in Figure 5 is the signature that the Universe was reionized after
decoupling. This is a subject rich in astrophysics, but for our
purposes it is important in that it provides another source for
scattering and hence detection of polarization. From the Wilkinson
Microwave Anisotropy Probe (WMAP) polarization data (11), one can
infer an optical depth of order 10\%, the fraction of photons
scattering in the reionized plasma somewhere in the region of $z =
10$. This new scattering source can be used to detect the primordial
gravity waves. The signature will show up at very low $l$ values,
corresponding to the horizon scale at reionization. Figure 6 shows
that the region $l = 4-8$ should have substantial effects from gravity
waves. Most likely, the only means of detecting such a signal is from
space, and even from there it will be very difficult. 

The polarization anisotropies for this very low $l$ region are
comparable to what is expected from the surface of last scattering ($l
\approx 100$). There are disadvantages to each signature. At the
lowest $l$ values, galactic foregrounds are more severe, there are
fewer modes in which to make a detection, and systematic errors are
likely greater. At the higher values, there is a foreground that
arises from E modes turning into B modes through gravitational lensing
(the topic of the next section). Clearly, it will be important to
detect the two signatures with the right relative strengths at these
two very different scales.
\subsubsection{Lensing of the CMB.} 
Both the temperature and polarization fields will be slightly
distorted (lensed) when passing collapsing structures in the late
Universe. The bending of light means that one is not looking (on the
last scattering surface) where one thinks. Although lensing will
affect both the polarization and T fields, its largest effect is on
the B field, where it shifts power from E to B. Gravitational
distortions, although preserving brightness, do not preserve the E and
B nature of the polarization patterns. 

Figure 6 also shows the expected power spectrum of these lensed B
modes. Because this power is sourced by the E modes, it roughly
follows their shape, but with $\Delta T$ suppressed by a factor of
20. The peak structure in the E modes is smoothed, as the structures
doing the lensing are degree scale themselves. Owing to the coherence
of the lensing potential for these modes, there is more information
than just the power spectrum, and work is ongoing to characterize the
expected cross correlation between different multipole bands. This
signal should be detectable in next-generation polarization
experiments. For our purposes, the most interesting aspect of this
lensing is the handle it can potentially give on the masses of the
neutrinos, as more massive neutrinos limit the collapse of matter
along the CMB trajectories. All other parameters held fixed, there is
roughly a factor-of-two change in the magnitude of the B signal for a
1-eV change in the mean neutrino mass.
\subsubsection{CMB scattering since reionization.} 
At very small angular scales -- $l$ values of a few thousand, way beyond
where the acoustic oscillations are damped -- there are additional
effects on the power spectra that result from the scattering of CMB
photons from electrons after the epoch of reionization, including
scattering from gas heated from falling deep in the potential wells of
Galaxy clusters (the Sunyaev- Zel'dovich, or SZ, effect). These
nonlinear effects are important as they can help in untangling (a)
when the first structures formed and (b) the role of dark energy.
\subsection{What We Learn from the CMB Power Spectrum} 
In this section, we show how the power spectrum information is used to
determine important aspects of the Universe. This is normally known as
parameter estimation, where the parameters are those that define our
cosmology. The observable power spectrum is a function of at least 11
such basic parameters. As we discuss below, some are better
constrained than others. 

First, there are four parameters that characterize the primordial
scalar and tensor fluctuation spectra before the acoustic
oscillations, each of which is assumed to follow a power law in wave
number. These four are the normalization of the scalar fluctuations
($A_s$), the ratio of tensor to scalar fluctuations $r$, and the
spectral indices for both (historically denoted with $n_s - 1$ and
$n_t$). Second, there is one equation-of-state parameter ($w$) that is
the ratio of the pressure of the dark energy to its energy density,
and one parameter that gives the optical depth ($\tau$) from the epoch
of reionization. Finally, there are five parameters that characterize
the present Universe: its rate of expansion (Hubble constant, with
$H_0 = h\cdot 100$ km s$^{-1}$ Mpc$^{-1}$), its curvature ($\Omega_k$), and
its composition (baryon density, matter density, and dark energy
density). The latter three are described in terms of energy densities
with respect to the critical density normalized to the present epoch:
$\omega_b = \Omega_b h^2, \omega_m = \Omega_m h^2$, and
$\omega_\Lambda = \Omega_\Lambda h^2$. Just 10 of these are
independent as $\Omega_m + \Omega_\Lambda + \Omega_k = 1$.

Even though the CMB data set itself consists of hundreds of
measurements, they are not sufficiently orthogonal with respect to the
10 independent parameters for each to be determined independently;
there are significant degeneracies. Hence, it is necessary to make
assumptions that constrain the values of those parameters upon which
the data have little leverage. In some cases, such prior assumptions
(priors) can have large effects on the other parameters, and there is
as yet no standard means of reporting results. 

Several teams have done analyses [WMAP (11, 12), CBI (13), Boomerang
(14), see also Reference 15]. Here we first discuss the leverage that
the CMB power spectra have on the cosmological parameters. Then we
give a flavor for the analyses, together with representative results. We
consider analyses, done by the several teams, with just the six most
important parameters: $\omega_b , \omega_m, A_s, n_s, \tau$ , and $h$,
where the other five are held fixed. For this discussion we are guided
by Reference 12. 

Completely within CMB data, there is a geometrical degeneracy between
$\Omega_k$, a contribution to the energy density from the curvature of
space, and $\Omega_m$. However, taking a very weak prior of $h > 0.5$,
the WMAP team, using just their first-year data, determined that
$\Omega_k = 0.03 \pm 0.03$, that is, no evidence for curvature.  We
assume $\Omega_k = 0$ unless otherwise noted. This conclusion has
gotten stronger with the three-year WMAP data together with other CMB
results, and it is a prediction of the inflationary
scenario. Nevertheless, we emphasize that it is an open experimental
issue.
\subsubsection{The geometry of the Universe.} 
The position of the first acoustic peak reveals that the Universe is
flat or nearly so. As we describe above, the generation of acoustic
peaks is governed by the (comoving) sound horizon at decoupling, $r_s$
(i.e., the greatest distance a density wave in the plasma could
traverse, scaled to today's Universe). The sound horizon depends on
$\omega_m, \omega_b$, and the radiation density, but not on $H_0,
\Omega_k, \omega_\Lambda$, or the spectral tilt $n_s$. The peak
positions versus angular multipole are then determined by $\Theta_A =
r_sd^{-1}_A$ , where the quantity $d_A$, the angular diameter
distance, is the distance that properly takes into account the
expansion history of the Universe between decoupling and today so that
when $d_A$ is multiplied by an observed angle, the result is the
feature size at the time of decoupling. In a nonexpanding Universe,
this would simply be the physical distance. The expression depends on
the (evolution of the) content of the Universe. For a flat Universe,
we have
\begin{equation}d_A = \int^{z_{dec}}_0 \frac{H_0^{-1} dz}{\sqrt{\Omega_r (1 + z)^4 + \Omega_m
(1 + z)^3 + \Omega_\Lambda}}.\end{equation} In this expression, $\Omega_r$ indicates
the (well-known) radiation density, and the dilutions of the different
components with redshift $z$, between decoupling and the present,
enter explicitly. 
\subsubsection{Fitting for spectral tilt, matter, and baryon content.} 
It is easy to see how one in principle determines spectral tilt. If
one knew all the other parameters, then the tilt would be found from
the slope of the power spectrum after the removal of the other
contributions.  However, there is clearly a coupling to other
parameters. Experiments with a very fine angular resolution will
determine the power spectrum at very high $l$ values, thereby improving
the measurement of the tilt. 

Here we discuss the primary dependences of the acoustic peak heights
on $\omega_m$ and $\omega_b$ . Increasing $\omega_m$ decreases the
peak heights. With greater matter density, the era of equality is
pushed to earlier redshifts, allowing the dark matter more time to
form deeper potential wells. When the baryons fall into these wells,
their mass has less effect on the development of the potential so that
the escaping photons are less redshifted than they would be, yielding
a smaller temperature contrast. As to $\omega_b$, increasing it
decreases the second peak but enhances that of the third because the
inertia in the photon-baryon fluid is increased, leading to hotter
compressions and cooler rarefactions (16). 

The peak-height ratios give the three parameters $n_s, \omega_m$, and
$\omega_b$, with a precision just short of that from a full analysis
of the power spectrum (discussed in Section 3.4.4). Following WMAP, we
define the ratio of the second to the first peak by $H^{TT}_2$, the
ratio of the third to the second peak by $H^{TT}_3$, and the ratio of
the first to the second peak in the polarization-temperature
cross-correlation power spectrum by $H^{TE}_2$. Table 1 shows how the
errors in these ratios propagate into parameter errors. We see that
all the ratios depend strongly on $n_s$, and that the ratio of the
first two peaks depends strongly on $\omega_b$ but is also influenced
by $\omega_m$. For $H^{TT}_3$, the relative dependences on $\omega_b$
and $\omega_m$ are reversed. Finally, the baryon density has little
influence on the ratio of the TE peaks. However, increasing $\omega_m$
deepens potential wells, increasing fluid velocities and the heights
of all polarization peaks. 

\begin{table}[htdp]
\begin{center}
\caption[Peak and Parameter Errors] {Matrix of how errors in the peak
 ratios (defined in text) relate to the parameter errors.}
\begin{tabular}{cccc}
\hline \hline
&$ \Delta n_s $& $ \frac{\Delta \omega_b}{\omega_b}$ &  $\frac{\Delta \omega_m}{\omega_m}$
\\
\hline $ \Delta  H_2^{TT}/H_2^{TT}$  & $0.88$ & $-0.67$ & $0.039$ \\
$\Delta  H_3^{TT}/ H_3^{TT}$ & $1.28$ & $-0.39$ & $0.46$ \\
$\Delta  H_2^{TE}/H_2^{TE}$ & $-0.66$& $0.095$ & $0.45$ \\
\hline
\end{tabular}
\end{center}
\end{table}

Table 2 lists the results from six-parameter fits to the power
spectrum from several combinations of CMB data with and without
complementary data from other sectors. The table includes results from
Reference 14, which included most CMB data available at time of
publication, and from even more recent analyses by WMAP (8).

\begin{center}
\begin{table}
\begin{tabular}{cccccc}
\hline \hline Symbol & WMAP1& WMAP3 & WMAP3       & CMB   & WMAP3 \\
                     &      &       & +other CMB  & + LSS & + SDSS\\
\hline  $\Omega_bh^2$ & $0.024 \pm 0.001$ & $0.02229 \pm 0.00073$ &$0.02232 \pm 0.00074$& $0.0226^{+0.0009}_{-0.0008}$ & $0.02230_{-0.00070}^{+0.00071}$\\
$\Omega_mh^2$ & $0.14 \pm 0.02$ & $0.1277_{-0.0079}^{+0.0080}$ &$0.1260\pm 0.0081$ & $0.143 \pm 0.005$ &$0.1327_{-0.0064}^{+0.0063}$\\
$h$ & $0.72 \pm 0.05$ & $0.732^{+0.031}_{-0.032}$ & $0.739^{+0.033}_{-0.032}$ & $0.695^{+0.025}_{-0.023}$ &$0.710\pm 0.026$\\
$\tau$ & $0.166_{-0.071}^{+0.076}$ & $0.089 \pm 0.030$ &$0.088_{-0.032}^{+0.031}$& $0.101_{-0.044}^{+0.051}$ & $0.080^{+0.029}_{-0.030}$\\
$n_s$ & $0.99 \pm 0.04$  & $ 0.958 \pm 0.016$ & $0.951 \pm 0.016$ & $0.95 \pm 0.02$ & $0.948_{-0.015}^{+0.016}$\\
\multicolumn{6}{l}{$\Omega_bh^2$:Baryon density, $\Omega_mh^2$: Matter density, $h$: Hubble parameter,} \\
\multicolumn{6}{l}{$\tau$: Optical Depth, $n_s$: Spectral index} \\
\hline
\end{tabular}
\caption{Results from six-parameter fits to CMB data, assuming a flat
Universe and not showing the scalar amplitude $A_s$ . Shown are
results from first-year WMAP data, three-year WMAP data, and WMAP data
combined with the bolometric experiments ACBAR and Boomerang. Fits
using data from CBI and VSA (using coherent amplifiers) were also
made, with consistent results. Also shown are results using LSS data
with CMB data available in 2003, and from adding LSS data [from the
Sloan Digital Sky Survey (SDSS)] to the WMAP3 data set (11). See
Section 1.11 for appropriate references.}
\end{table}
\end{center}
\subsection{Discussion of Cosmological Parameters}
The overall conclusions from the analysis of the peak structure are
not dramatically different from those drawn from a collection of
earlier ground- and balloon-based experiments. Still, WMAP's first
data release put the reigning cosmological model on much stronger
footing. Few experiments claimed systematic errors on the overall
amplitude of their TT measurements less than 10\%; WMAP's errors are
less than 0.5\%. The overall amplitude is strongly affected by the
reionization. With full-sky coverage, WMAP determined the power
spectrum in individual $l$ bins with negligible correlations. Now with
the WMAP three-year data, results from higher-resolution experiments,
and results on EE polarization, we are learning even more. 

Remarkably, CMB data confirm the baryon density deduced from Big Bang
nucleosynthesis, from processes occurring at approximately 1~s after
the Big Bang: $\Omega_b h^2 = 0.0205 \pm 0.0035$. The determination of
the nonzero density of dark matter at approximately 300,000 years
reinforces the substantial evidence for dark matter in the nearby
Universe. Finally, the flat geometry confirms the earlier (supernova)
evidence of a dark energy component. 

With temperature data alone, there is a significant degeneracy in
parameter space, which becomes apparent when one realizes that there
are just five key features in the power spectrum (at least with
today's precision) to which one is fitting six parameters: the heights
of three peaks, the location of the first peak, and the anisotropy on
very large scales. The degeneracy can be understood as follows. The
peak heights are normalized by the combination $A_s e^{-2\tau}$. Thus,
both these parameters can increase in a way that leaves the peak
heights unchanged, increasing the power on scales larger than the
horizon at reionization. Increasing $n_s$ can restore the balance but
can also decrease the second peak. That peak can be brought back up by
decreasing $\omega_b$. WMAP broke this degeneracy in its first-year
release with a prior requiring $\tau < 0.3$. The new EE data from
WMAP, in their sensitivity to reionization, break this degeneracy
without the need for a prior.

Table 2 shows that the Hubble constant has been robust and in good
agreement with determinations from Galaxy surveys. With better
measurements of the peaks, the baryon and matter densities have moved
systematically, but within error. The optical depth has decreased
significantly and is now based upon the EE, rather than TE, power in
the lowest $l$ range. This change is coupled to a large change in the
scalar amplitude. Finally, evidence for a spectral tilt ($n_s \ne 1$)
is becoming more significant. As this is predicted by the simplest of
inflationary scenarios, it is important and definitely worth
watching. 

The first-year WMAP data confirmed the COBE observation of
unexpectedly low power in the lowest multipoles. The WMAP team
reported this effect to be more significant than a statistical
fluctuation, and lively literature on the subject followed. It is it
clear that the quadrupole has little power and appears to be aligned
with the octopole. However, the situation is unclear in that the
quadrupole lines up reasonably well with the Galaxy itself, and there
is concern that the cut on the WMAP data to remove the Galaxy then
reduced the inherent quadrupole power. The anomaly has been reduced
with the three-year data release, with improvements to the analysis,
particularly in the lowest multipoles. 

Table 2 also gives results from fitting CMB data with data from other
cosmological probes, in particular large-scale structure (LSS) data in
the form of three-dimensional Galaxy power spectra (the third
dimension is redshift). Such spectra extend the lever arm in $k$
space, allowing a more incisive determination of any possible spectral
tilt, $n_s$. However, there are potential biases with the Galaxy
data. In particular, the galaxies may not be faithful tracers of the
dark matter density. Already before the three-year WMAP data release,
including LSS data with CMB data favored an optical depth closer to
its current value and provided evidence of spectral tilt. With
three-year WMAP data and the Sloan Digital Sky Survey Galaxy survey
data, the significance of a nonzero tilt is near the 3$\sigma$
level. This is a vigorously debated topic. There are other LSS surveys
that give similar and nearly consistent results, yet the systematic
understanding is not at the level where combining all such surveys
makes sense.
\subsubsection{Beyond the six basic parameters.}
With the LSS data, one can obtain information on other parameters that
were held fixed. In particular, relaxing the constraint on $\Omega_k$,
one finds consistency with a flat Universe to the level of approximately
0.04 (with CMB data alone) and 0.02 (using LSS data) (see Reference
14). Using WMAP and other surveys, constraints as low as 0.015 are
obtained with some sets, giving slight indications for a closed
Universe ($\Omega_k < 0$). 

There is sensitivity to the fraction of the dark matter that resides
in neutrinos: $f_\nu$. The neutrino number density (in the standard
cosmological model) is well known; a mean neutrino mass of 0.05~eV
corresponds to $\Omega_\nu$ of approximately 0.001. The current limits
are $\bar{M_\nu} < 1~$eV from the CMB alone and $\bar{M_\nu} < 0.4~$eV
when including Galaxy power spectra (14). 

One can also extract information about the dark energy
equation-of-state parameter $w$. If dark energy is Einstein's
cosmological constant, then $w \equiv -1$. Because $w$ affects the expansion
history of the Universe at late times, the associated effects on power
spectra then give a measure of $w$. Using all available CMB data,
Reference 14 finds $w = -0.86^{+0.35}_{-0.36}$. However, by including
both the Galaxy power spectra and SN1A data, the stronger constraint
$w = -0.94^{+0.093}_{-0.097}$ is derived. WMAP, using its own data and
another collection of LSS data together with supernova data, finds $w
= -1.08 \pm 0.12$, where in this fit they also let $\Omega_k$
float. 

Finally, we want to mention a new effect, even if outside the domain
of the CMB baryon oscillations. In principle, one should be able to see
the same kind of acoustic oscillations in baryons (galaxies) seen so
prominently in the radiation field. If so, this will provide another
powerful measure of the effects of dark energy at late times,
specifically the time when its fraction is growing and its effects in
curtailing structure formation are the largest. This effect has
recently been seen (17) at the level of 3.4$\sigma$, and new
experiments to study this far more precisely are being proposed. This
is an excellent example of how rapidly the field of observational
cosmology is developing. In the wonderful textbook Modern Cosmology by
Scott Dodelson (2, p. 209), Dodelson states that this phenomenon would
only be barely (if at all) detectable.  

Before turning to a discussion of the problem of astrophysical
foregrounds, we mention that currently the utility of ever more
precise cosmological-parameter determination is, like in particle
physics, not that we can compare such values with theory but rather
that we can either uncover inconsistencies in our modeling of the
physics of the Universe or gain ever more confidence in such modeling.
\section{FOREGROUNDS}
 Until now, we have introduced the features of the CMB, enticing the
 reader with its promises of fascinating insights to the very early
 Universe. Now we turn our attention toward the challenge of actually
 studying the CMB, as its retrieval is not at all an easy
 endeavor. Instrumental noise and imperfections could compromise
 measurements of the tiny signals (see Section 3). Even with an ideal
 receiver, various astrophysical or atmospheric foregrounds could
 contaminate or even suppress the CMB signal. In this section we first
 give an overview of the relevant foregrounds, then describe the
 options for foreground removal and estimate their
 impact. 
\subsection{Overview} 
\begin{figure}[h]
\begin{center}
\includegraphics[angle=270, width=6in]{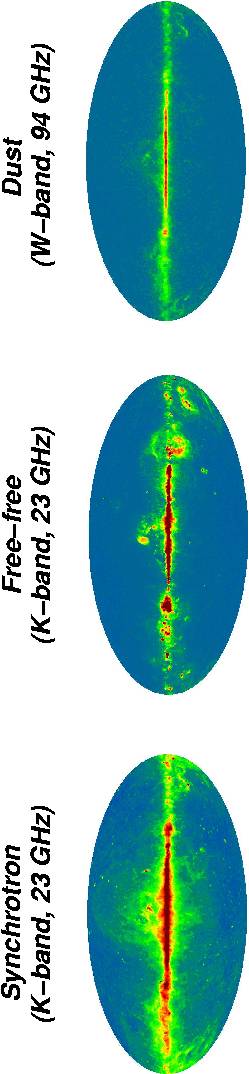}
\caption{Unpolarized foreground maps in Galactic coordinates, derived
from WMAP. Each map is shown at the WMAP frequency band in which that
foreground is dominant. The color scale for the temperature is linear,
with maxima set at approximately 5~mK for K-band and 2.5~mK
forW-band. Images courtesy of the WMAP science team.}
\end{center}
\end{figure}
One may be tempted to observe the CMB at its maximum, approximately
150-200 GHz. However, atmospheric, galactic, or extragalactic
foregrounds, which have their own dependences on frequency and angular
scale, may dominate the total signal, so the maximum may not be the
best choice. 

The main astrophysical foregrounds come from our own Galaxy, from
three distinct mechanisms: synchrotron radiation; radiation from
electron-ion scattering, usually referred to as free-free emission;
and dust emission. Figure~7 displays full-sky intensity maps for the
main foreground components as derived from WMAP data at microwave
frequencies where the bright Galaxy is clearly dominating the
pictures. Each component is shown for the WMAP frequency channel where
it is dominant.

Figure 8 compares the expected CMB signal as a function of frequency
to the rms of WMAP foreground maps on an angular scale of
1$^\circ$. The ordinate axis records antenna temperature (see Section
3.2.1). An optimal observing frequency range with the highest ratio of
CMB to foreground signal is in the region around 70 GHz (often termed
the cosmological window).

Much less is known about the polarization of foregrounds. Information
is extrapolated mostly from very low or very high frequencies or from
surveys of small patches. Figure 8b shows an analog figure for the
polarization fluctuations as estimated from WMAP three-year data on an
angular scale of approximately $2^\circ (l = 90)$, where the signal
from gravitational waves is maximal. The dust estimate has some
limitations because the WMAP frequency channels do not extend to the
high frequencies where the dust is expected to dominate the
foregrounds. 
\begin{figure}
\begin{center}
\includegraphics[width=2.6in]{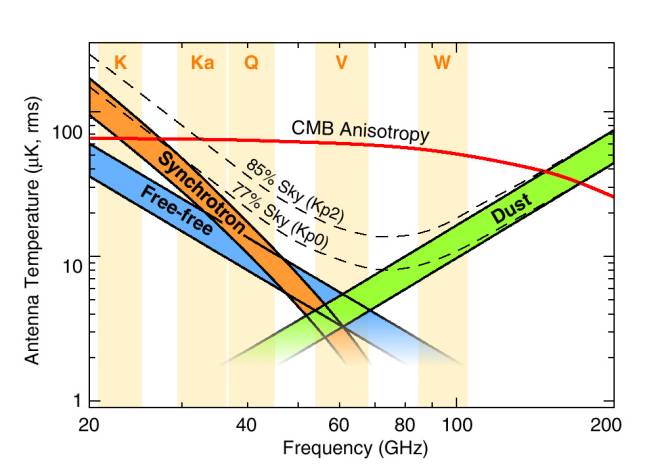}
\includegraphics[width=2.6in]{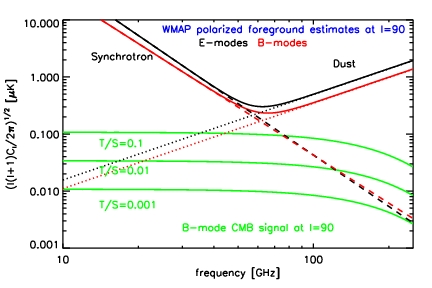}
\caption{Frequency dependence of foregrounds recorded in antenna
temperature. (a) The rms on angular scales of 1æ for the unpolarized
CMB compared with that from foregrounds extracted from the WMAP data
(18). The WMAP frequency bands (K, Ka, Q, V, W) are overlaid as light
bands. These plots are for nearly full sky; the total foregrounds are
shown as dashed lines for two different sky cuts. Figure courtesy of
the WMAP science team. (b) A similar plot of the expected polarization
level of foregrounds at $l = 90$ in comparison with that from
primordial B modes (which peak around $l = 90$) for different values
of r following formula 25 in Reference 19. Again, these estimates are
for observations covering most of the sky.}  
\end{center}
\end{figure}

The expected B-mode signal is smaller than the estimated foreground
signal even for $r = 0.1$. However, almost the full sky was used for
the estimate, whereas recent studies (20, 21) using lower-frequency
data and WMAP data indicate that the polarization of synchrotron
radiation on selected clean patches can be significantly
smaller. Thus, the optimal frequency window will shift depending on
which region is observed. After discussing possible foreground effects
from Earth's atmosphere, we briefly review what is known about the
dominant sources of galactic and extragalactic foregrounds.
\subsubsection{Atmospheric effects.} 
The atmosphere absorbs short-wavelength radiation, but fortunately has
transmission windows in the range of visible light and microwave
radiation. Absorption lines from oxygen (around 60 and 120 GHz) and
water vapor (20 and 180 GHz) limit the access to the microwave sky,
and, in particular, clouds and high water vapor can compromise
ground-based observations. Thermal emission from the atmosphere can
add significantly to the observed signal for ground-based experiments
(depending on the observing site and the frequency, from 1-40 K) and,
together with the instrumental noise and/or thermal emission from warm
optical components, can make for the major part of the detected power
(see also Section 3.2.1). The observing strategy needs to be designed
in a way that allows a proper removal of the varying atmospheric
contribution without a big impact on the signal extraction (see also
Sections 3.3 and 3.4.1). 

Although thermal emission from the atmosphere is unpolarized, the
Zeeman splitting of oxygen lines in Earth's magnetic field leads to
polarized emission, which is dominantly circularly polarized. Although
the CMB is not expected to be circularly polarized, Hanany \&
Rosenkranz (22) showed that for large angular scales, $l \approx 1$, a
0.01\% circular-to-linear polarization conversion in the instrument
could produce a signal more than a factor of two higher than the
expected gravitational wave B-mode signal if $r$ were small, that is,
if $r = 0.01$. 

In addition, backscattering of thermal radiation from Earth's surface
from ice crystal clouds in the upper troposphere may give signals on
the order of micro-Kelvin size (23), again larger than the expected
B-mode signal. Although the polarized signal from oxygen splitting
would be fixed in Earth's reference frame, and thus could be separated
from the CMB, the signal from such ice clouds would reflect the
varying inhomogeneous cloud distribution and thus be hard to remove.
\subsubsection{Galactic synchrotron radiation.} 
Synchrotron radiation is something familiar to particle physicists,
mostly from storage rings where some of the energy meant to boost the
particle's energy will be radiated away. The same effect takes place
in galactic accelerators, with cosmic-ray electrons passing through
the galactic magnetic field. In contrast to the particle physics case,
where electrons of energies of a few GeV pass magnetic elds of up to a
few 1000~G, we are dealing here with electrons in a galactic field of
only a few micro-Gauss. 

This component of the foreground radiation is dominant at frequencies
below 70 GHz, and its intensity characteristics have been studied at
frequencies up to 20 GHz. The frequency and angular dependence both
follow power laws $T\propto \nu^{-\beta}$ , with a position- and
frequency-dependent exponent that varies between 2 and
3. 

Theoretically, a high degree of sychrotron polarization ($>75\%$) is
expected, but low-frequency data imply much lower values. However, at
low frequencies, Faraday rotation where light traversing a magnetized
medium has its left and right circular polarized components travel at
different speeds reduces the polarization.
\subsubsection{Galactic dust.} 
Interstellar dust emits mainly in the far infrared and thus becomes
relevant for high frequencies ($\nu > 100$~GHz). The grain size and
dust temperature determine the properties of the radiation, where the
intensity follows a power law $T\propto T_0 \nu^\beta$, with the
spectral index $\beta \approx 2$ and with both $T_0$ and $\beta$
varying over the sky. Using far-infrared data from COBE, Finkbeiner et
al. (24) (FDS) provided a model for the dust emission consisting of
two components of different temperature and emissivity ($T =
9.4/16$~K, $\beta = $1.67/2.7). 

There are also indications for another component in the dust emission,
as seen through cross correlation of the CMB and far-infrared
data. Its spectral index is consistent with free-free emission, but it
is spatially correlated with dust. This anomalous dust contribution
could derive from spinning dust grains. However, current data do not
provide a conclusive picture, and additional data in the 5-15-GHz
range are needed to better understand this component (25). In 2003,
the balloon-borne experiment ARCHEOPS reported 5\% to 20\%
polarization of the submillimeter diffuse galactic dust emission,
providing the first large coverage maps of polarized galactic
submillimeter emission at 13' resolution (26). More recently, they
also published submillimeter polarization limits at large angular
scales, which when extrapolated to 100~GHz are still much larger than
the expected gravitational wave signal for $r = 0.3$ (27).
\subsubsection{Free-free emission.} 
Electron-ion scattering leads to radiation that is, in this context,
termed free-free emission, whereas in the high-energy lab, it is
better known as bremsstrahlung. This component does not dominate the
foregrounds at any radio frequency. Sky maps of free-free emission can
be approximated using measurements of the H$\alpha$ emission (from the
hydrogen transition from n = 3 to n = 2), which traces the ionized
medium. The thermal free-free emission follows a power law $T \propto
\nu^{-\beta}$, where $\beta \approx 2$. This foreground is not
polarized. 
\subsubsection{Point sources.} 
Known extragalactic point sources are a well-localized contaminant and
easily removable. However, the contribution from unresolved point
sources can severely affect measurements: for example, the recent
discussion of their impact on the determination of ns fromWMAPdata
(28). Point sources impact CMB measurements mostly at high angular
scales and low frequencies. For low frequencies, their contribution
may still be larger than the signal expected from gravitational
waves.
\subsection{Foreground Removal} 
Understanding and removing foregrounds are most critical for the tiny
polarization signals. The different frequency dependences of the CMB
and galactic foregrounds provide a good handle for foreground removal
using multifrequency measurements. 

For the polarization analysis, methods where little or no prior
information is required are the most useful for now. A promising
strategy is the Independent Component Analysis, which has already been
applied to several CMB temperature data sets (including COBE, BEAST,
and WMAP) and for which formalism has also been developed to cope with
polarization data. The foreground and CMB signals are assumed to be
statistically independent, with at least one foreground component
being non-Gaussian. Then the maximization of a specific measure of
entropy is used to disentangle the independent components. Stivoli et
al. (29) demonstrated a successful cleaning of foregrounds using
simulated data. Verde et al. (30) estimated the impact of foregrounds
independent of removal strategy, considering different degrees of
effectiveness in cleaning. A 1\% level of residual foregrounds, in
their power spectrum, was found to be necessary to obtain a 3$\sigma$
detection of $r = 0.01$ from the ground. 

Because all current studies rely on untested assumptions about
foregrounds, they need to be justified with more data. Moreover, none
of the studies to date takes into account the impact of foregrounds in
the presence of lensing and instrumental systematics. Work is needed
on both the experimental and theoretical side to obtain a more
realistic picture of the foregrounds and their impact.
\section{METHODS OF DETECTION} 
We have argued that in the patterns of the CMB lies greatness; here we
outline the essential ingredients for measuring CMB anisotropies. The
fundamental elements for detecting microwave emission from the
celestial sphere are optics and receivers. The optics comprises
telescopes and additional optical elements that couple light into the
receivers. The receivers transduce the intensity of the incoming
microwave radiation into voltages that can be digitized and
stored. Two other CMB experiment requirements are fidelity control
(calibration and rejection of spurious signals) and optimized
strategies for scanning the telescope beams across the rotating
sky. Below, after a general introduction to the problem, we elaborate
on these topics and culminate with an overview of data analysis
techniques. 
\subsection{The CMB Experiment Basics} 
All CMB experiments share certain characteristics. Some main optical
element determines the resolution of the experiment. This main optical
element may be a reflecting telescope with a single parabolic mirror,
or one with two or more mirrors; it may be a refracting telescope
using dielectric lenses; it may be an array of mirrors configured as an
interferometer; it may be just a horn antenna \footnote{A horn antenna
is waveguide flared to the appropriate aperture for the desired
resolution; these were used in the COBE satellite instrument that made
the first detection of CMB anisotropy.}. In most cases, additional
optical elements are required to bring the light to the
receiver. Examples include Dewar windows, lenses, filters, polarization
modulators, and feedhorns (which are horn antennas used to collect
light from telescopes). Typically these coupling optics are small
enough that they can be maintained at cryogenic temperatures to reduce
their thermal emission and lossiness. 

The low-noise receivers are nearly always cryogenic and divide into
two types, described below. Spatial modulation of the CMB signal on
timescales of less than one minute is critical to avoid slow drifts in
the responsivity of the receivers, and may proceed by movement of the
entire optical system, or by moving some of its components while
others remain fixed. Large ground screens surround most experiments to
shield the receivers from the 300-K radiation from Earth. Typically,
the thermal environment of the experiment must be well regulated for
stability of the receiver responsivity and to avoid confusion of
diurnal effects from the environment with the daily rotation of the
celestial signal. Earth-bound experiments suffer the excess noise from
the atmosphere, as well as its attenuation of the signal, and must
contend with 2$\pi$ of the 300-K radiation. Balloon-borne experiments
suffer less atmosphere, but must be shielded from the balloon's
thermal radiation and typically have limited lifetimes (1-20
days). Long flights usually require constant shielding from the Sun
during the long austral summer day. Space missions have multiple
advantages: no atmosphere, Earth filling a much less solid angle, a
very stable thermal environment, and a longer lifetime than current
balloon missions.
\subsection{The Detection Techniques} 
Although to fully describe the CMB anisotropies requires their spatial
power spectra (which happily are not white), a useful
order-of-magnitude number is that the rms of the CMB sky when
convolved to 10$^\circ$ scales is approximately 30 $\mu$K, and
approximately 70~$\mu$K for 0.7$^\circ$ scales (the first acoustic
peak). This rms of the CMB temperature is some 20 ppm of the 2.7-K
background, the polarization E modes are 20 times lower, and the
primordial B-mode rms is predicted to be 50 ppb or less. 

A microwave receiver measures one or more of the Stokes parameters of
the radiation incident on it. Two classes of low-noise receivers may
be identified: coherent receivers, in which phase-preserving
amplification of the incident field precedes detection of its
intensity, and incoherent receivers, in which direct measurement of
the intensity of the incident field is performed. 

In coherent receivers, the incident field is piped around transmission
lines as a time-varying voltage. That voltage is amplified in
transistor amplifiers, and then the signal is eventually detected when
it passes through a nonlinear element (such as a diode) with an output
proportional to the square of the incident field strength. The critical
element in the coherent receiver is not the detector but typically the
transistor amplifier, which must be a low-noise amplifier. In cases
where transistor amplifiers are not available at high enough
frequencies, the first and most critical element in the receiver is a
low-noise mixer, which converts the frequency of the radiation to
lower frequencies, where low-noise amplifiers are available. 

For the CMB, the most widely used incoherent detector to date is the
bolometer. A bolometer records the intensity of incident radiation by
measuring the temperature rise of an isolated absorber of the
radiation. A promising effort is underway to develop receivers in
which the bolometers are coupled to transmission lines, where they can
serve as the very low-noise detectors in what otherwise looks like a
coherent receiver.
\subsubsection{Calibration, Kelvins, and system temperature.} 
A microwave receiver outputs a voltage proportional to the intensity
$I$ of the incident radiation over some effective bandwidth $\Delta
\nu$ centered on frequency $\nu_0$. The output is calibrated in
temperature units through observation of blackbody sources. The
polarization anisotropies of the CMB are also described in temperature
units. This follows because the Stokes parameters $Q$ and $U$ have the
same units as the intensity $I$. There is a factor of two to keep
track of: The usual definition of $I$ sums the intensities from two
orthogonal polarization modes. Note that the antenna temperature $T_A$
is defined by the approximation $I \propto T_A$. Only in the
Rayleigh-Jeans regime of a blackbody does $T_A \approx T$, but it is a
convenient measure for comparing the effects of various foregrounds
and other contaminants. 

Microwave receivers are sensitive to the total intensity of the
incident radiation over the bandpass. The incident electric field can
be considered as a sum of incoherent (i.e., uncorrelated) sources,
each with intensity that can be associated to a temperature in the
Rayleigh-Jeans limit. Thus, we can define the system temperature to
describe the input power to the receiver:
\begin{equation}T_{sys} = T_{CMB} + T_{fg} + T_{atm} + T_{gnd} + T_{opt} + T_n,\end{equation} where
we have included terms for the CMB, foregrounds, atmosphere, 300-K
emission from the ground, emission from the warm optics, and receiver,
respectively.We do not note explicitly that the extra-atmospheric
signals are attenuated slightly as they pass through the
atmosphere. At good sites such as the Atacama Desert in Chile or the
South Pole, this effect is small for $\nu = 110$~GHz. We also neglect
absorption in the optics, although in bolometer systems, this effect
can be large. Note that when describing bolometer receivers, it is
more common to leave the sum in units of power, as we see below.
\subsubsection{Sensitivity and noise.} 
Imagine a CMB experiment that scans across a small enough region of
sky that the sky curvature may be neglected, recording the temperature
of each of N beam-sized patches of sky a single time into a vector
$d$. The error on each measurement is $\sigma_e$. Let us first take
the (unattainable) case where $\sigma_e \ll 1~$nK. In that case,
$\sigma_d$ measures the variance of the CMB itself. If the CMB power
spectrum were white with average level $\Delta T^2$, for example, the
variance could be crudely estimated as
\begin{equation} \sigma_d^2\approx \Delta T^2 \frac{\delta l}{l_c},\end{equation} 
where $l_c$ is some average in the region $\delta l$ between
($x_b)^{-1}$ and $(Nx_b)^{-1}$, with $x_b$ the diameter of the
beam-sized patch in radians. Typical numbers might give $\delta l/l_c
\approx 1$, and $\Delta T = 60\mu$~K $\approx \sigma_d$. Thus, for
$\sigma_e$ so small, one could estimate the Fourier modes (or the
$C_l$ themselves) directly and trace out details of the spectrum over
the range $\delta l$. 

To consider a more realistic case, we note that CMB receivers are
characterized according to their sensitivity $S$ in units of
K$\sqrt{s}$ . The variance $\sigma_T^2$ of a series of measurements,
each resulting from an integration time $\tau$, is then found by
$\sigma_T^2= S/\tau$. A typical value for the sensitivity of a single
receiver is 500~$\mu$K$\sqrt{s}$ , so that after 10 min on each patch,
one might attain $\sigma_e \approx 20 \mu$K and then detect the CMB
signal excess variance in the data. Typical variances in the CMB E
polarization are ($4~\mu$K$)^2$, so these experiments must integrate
for hundreds of hours and/or use hundreds of receivers. The tougher
sensitivity requirements for B modes are described in Section 4. 

In cases where the mean photon mode occupancy $n_0$ for the input
radiation is large, $n_0 = (e^x - 1)^{-1} \gg 1$ with $x = h \nu
/(kT_{sys})$, we can describe classically the sensitivity of an ideal
direct receiver of bandwidth $\Delta \nu$ with the Dicke equation: $S
= T_{sys}/\sqrt{\Delta \nu}$. We account for more complex receivers
below. Most coherent receivers operate in the limit $n_0 \gg 1$. In
cases where $n_0 \ll 1$ (low $T_{sys}$ and high frequency $\nu$), the
limitation to sensitivity in ideal bolometric receivers comes from
photon shot noise (counting statistics), and the sensitivity depends
on $T_{sys}$ in cases where the bolometer thermal noise is
negligible. We return to the noise in bolometers below. 

The sensitivity of the receiver captures its best features
succinctly. It is also instructive to look at the spectra of
postdetection signal noise from the receivers. The scanning of the
telescope translates the CMB anisotropy signal to variations in time
that lie atop the intrinsic postdetection noise. A typical noise power
spectrum is shown in Figure 9. This spectrum shows a white-noise level
at frequencies $f > 0.01$~Hz; the sensitivity is measured from the
white noise. The spectrum also shows characteristic low-frequency
noise approximately proportional to $1/f$, which can be parametrized
by the frequency $f_c$ where the $1/f$ and white-noise powers are
equal. Such $1/f$ noise is ubiquitous; the atmosphere itself has a
$1/f$ spectrum. In principal, this noise is quite serious, as it
contributes more and more to the variance with longer integration
times. In practice, experiments are designed to modulate the signal at
frequencies $f > f_c$. Usually the receiver includes a low-pass filter
that limits the bandwidth of the postdetection data stream to
$f_{Ny}$. Then these data are Nyquist sampled and digitized at a rate
of $2f_{Ny}$ to avoid aliasing high-frequency noise. Roughly speaking,
the postdetection bandwidth of interest is usually between 0.01-0.1 Hz
and 50-200 Hz.
\subsubsection{Coherent receivers.} 
A key advantage of coherent receivers is that they can be configured
so that their output is the correlation of two input signals. We
devote the bulk of our discussion here to this topic. 
\paragraph{Coherent-receiver noise spectra.} The noise properties of
amplifiers include both intrinsic output voltage fluctuations, present
in principle even in the absence of input signal \footnote{We remind
the reader that for the frequencies of interest, absence of input
signal is a difficult requirement because a resistive termination
emits thermal radiation, whereas a shorted input reflects back any
thermal radiation emitted from the amplifier input.} and characterized
by $T_n$, and fluctuations in the gain coefficient (the factor by 
which the input voltage is amplified).
\begin{figure}
\hspace*{-3cm}
\includegraphics[width=8in]{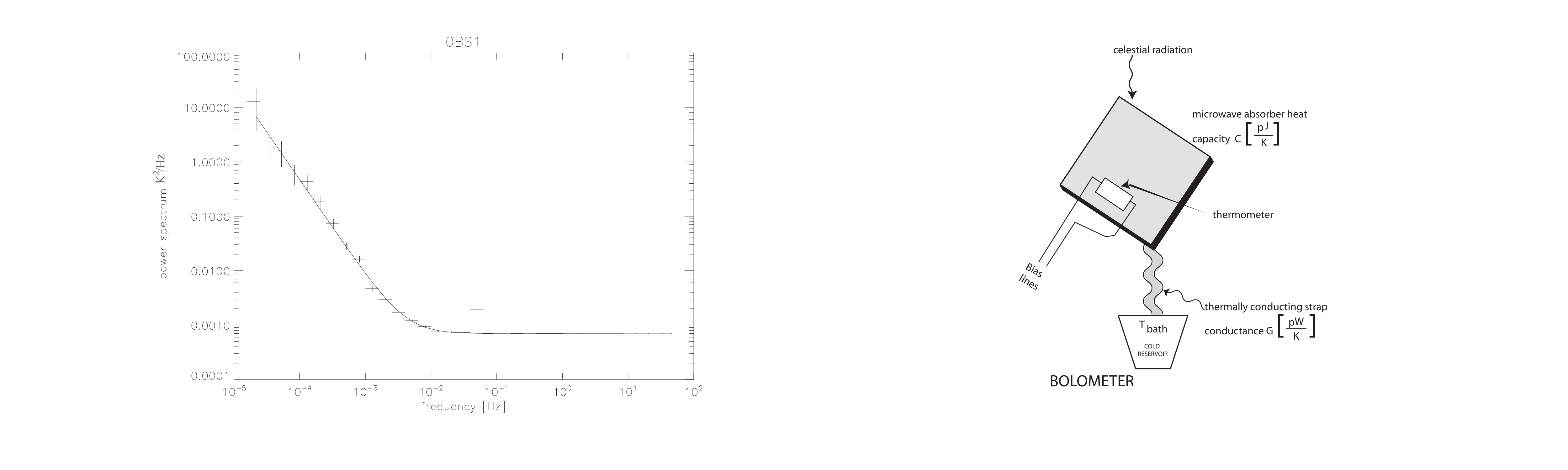}
 \caption{(Left) Raw postdetection noise power spectrum from a
correlation polarimeter during scans of the sky, illustrating three
interesting features: its white-noise level at frequency $f >
0.01$~Hz, a peak corresponding to the scan period of 21 s, and the
characteristic low-frequency noise with slope $1/f$. Figure courtesy
of C. Bischoff and the CAPMAP team (31). (Right) Sketch of a
bolometer, indicating that celestial radiation is absorbed in a
material with heat capacity $C$ coupled to a cold reservoir at
temperature $T_{bath}$ by a strap with low thermal conductance $G$. A
thermometer measures the power delivered to the absorber by recording
its temperature increase.}
\end{figure} 
An extension of the Dicke equation reads
\begin{equation} 
\delta T = a T_{sys} \sqrt{\frac{1}{\Delta \nu \tau} + \left (\frac{\delta g}{g}\right )^2}, 
\end{equation} 
where the coefficient $a$ depends on the exact configuration of the
receiver; $a = 1$ for a direct receiver in which the input is
amplified and then detected. For the correlation polarimeters
described below, $a = \sqrt{2}$.
\paragraph{Signal correlation.} 
Two signals are correlated if the time average of their product is
nonzero. In 1952, Ryle (32) introduced the concept of phase switching:
periodically introducing a half-wavelength phase lag into one of the
two signal lines, which changes the sign of the correlation
product. Then, one can sum the two inputs and square them in a
detector diode. The correlation product is now modulated at the
phase-switching rate and may be recovered. The correlation can also be
measured in analog correlators (multipliers) or by digitization of the
inputs, followed by multiplication.
\paragraph{Correlation receivers.} 
Correlation receivers have been used in several ways to measure CMB
signals. We describe two methods below. The advantages of correlation
include reduced sensitivity to $1/f$ variations in amplifiers (because
phase switching with diode switches can be effected at kilo-Hertz
rates), reduced sensitivity to gain fluctuations (because those
fluctuations multiply only the small correlated signal rather than the
large, common-mode intensity signal), and the ability to access all
four of the Stokes parameters that describe the full polarization
state of the radiation using only two inputs.
\paragraph{Interferometers.} 
Perhaps the best-known example of signal correlation in astronomy lies
in the interferometer, which uses the correlation of signals from two
spatially separated telescopes to measure Fourier modes of the
celestial radiation in a limited field of view. The interferometer's
receiver detects a slice of the interference pattern that arises when
the two input signals originate with phase coherence, much as the
screen in a Young's diffraction experiment does. The angular
resolution is determined by the spacing between the telescopes,
whereas the size of the individual telescopes limits the field of
view. The resolution of an interferometer may be described in terms of
its synthesized beamwidth, $\sigma_b$. Because the CMB is an extended
source, nearly all the Fourier modes in the field of view are needed
to fully characterize it. All the Fourier modes can be measured only
if the individual telescopes are close packed so that they fill all the
space required for an equivalent single dish of resolution $\sigma_b$. 

To recover all four Stokes parameters, the telescope optics usually
includes circular polarizers so that a given receiver amplifies either
the left ($L$) or right ($R$) circular polarization state. Both DASI
and CBI used quarter-wave plates in the optics to periodically reverse
the polarization state, $L \leftrightarrow R$, so that all possible
combinations of correlations ($LR, LL, RR,$ and $RL$) could be
measured from each pair of telescopes.
\paragraph{Correlation polarimeters.} 
In a correlation polarimeter, the incident signal enters through an
azimuthally symmetric feedhorn and is separated into two polarization
states (either two orthogonal linear polarization states or the $L$
and $R$ circular polarization states) before amplification in
high-electron-mobility-transistor amplifiers. After amplification, the
two polarization states are correlated so that the output is
proportional to a linear Stokes parameter: $V_{out} \propto \pm E_x
E_y \propto \pm U$, with the modulation provided by the phase
switch. Figure 9 shows the stability of a phase-switched correlation
polarimeter, switched at 4 kHz. The high-electron-mobility-transistor
amplifiers in this polarimeter have $1/f$ knees at 1~kHz and cannot be
used without this rapid modulation. The sensitivity of this
polarimeter is found from the Dicke equation with $a =\sqrt{2}$. If
$L$ and $R$ states are used, then the second linear Stokes parameter
can also be obtained by phase shifting $R$ by an extra $\pi/2$ and
then correlating it with $L$. The correlation can come about either
through direct multiplication of the two polarization states or
through the Ryle technique. The QUIET project uses the latter method,
in which $L$ and $R$ are sent into a compact module that mounts onto a
circuit board. The correlation $LR$ is constructed by differencing the
squares of the sum and difference terms $L + R$ and $L - R$; $U$ is
found by similar means. Phase switches at 4~kHz modulate the outputs,
which are read out on module pins; other module pins are used to bias
the amplifiers.
\subsubsection{Incoherent detectors.} 
A number of clever ideas beyond the scope of this review are being
pursued for tailoring bolometers to search for CMB polarization. One
recent great success was the advent of the polarization-sensitive
bolometer. Another avenue is using wafer-level silicon fabrication
techniques to produce arrays of hundreds of detectors at once,
sometimes coupling the detectors directly to planar antennas or
antenna arrays (antenna-coupled bolometers). Below we focus on the
rudiments of the transition edge sensor (TES) bolometer. Figure 9
depicts the critical elements of a bolometer. Celestial radiation
impinges on a microwave absorber with heat capacity $C$. The absorber
is connected to a cold reservoir at temperature $T_{bath}$ by a
thermal strap of heat conductance $G$. A fluctuation in the intensity
of the celestial radiation warms the absorber slightly. This
temperature change is recorded by a thermometer, typically a material
with a large-magnitude logarithmic derivative $d log R/d log T \equiv
\alpha$ of resistance with respect to temperature. 
\paragraph{Transition edge sensor bolometers.}
The TES is a superconductor maintained at its critical temperature
$T_c$. We describe the TES as being on (or in) the transition when its
resistance is between zero and the normal resistance $R_n$. For CMB
devices optimized to work with $T_{bath} \approx 300$~mK, a typical
critical temperature is $T_c \approx 400-500$~mK. The width of the
transition is typically a percent of $T_c$. Many TESs comprise a
bilayer: a thin layer of natural superconductor topped with normal
metal. The $T_c$ and $R_n$ of bilayers can be controlled by varying the two
thicknesses. 

The TES is operated in series with an ideal inductor $L$. To voltage
bias the TES at operating resistance $R_0$, a small shunt resistance
$R_{sh} \ll R_0$ is placed in parallel with the TES/inductor
combination and fed with a bias current. Fluctuations in $R_0$ cause
current fluctuations through the inductor. Flux through the inductor
is linked to a superconducting quantum interference device (SQUID),
which serves as a current amplifier. To prevent the SQUID current
noise from contributing significantly, the operating resistance $R_0$
is kept low: $R_0 < 1 \Omega$ .

The voltage bias causes the TES resistor to dissipate heat. Rough
order-of-magnitude values for TES bolometers used for the CMB may be
$G \approx 30$~pWK$^{-1}$ and $C \approx 0.3$~pJ K$^{-1}$. The thermal
timescale for the bolometer to change temperature in response to a
fluctuation in the photon power is $\tau_{th} = C/G \approx 10$~ms,
corresponding to a one-pole low-pass filter at $f = (2\pi\tau)^{-1}
\approx 15$~Hz. Conceptually, we note that when a fluctuation in the
photon noise warms the bolometer, the TES resistance increases, as
$\alpha >0$. Because the TES is voltage biased, the increase in
resistance lowers the Joule heating power $V^2/R$ flowing into the
bolometer. These electrical changes can occur more quickly than the
thermal timescale $\tau_{th}$. This is the phenomenon of
electrothermal feedback, which speeds up the TES bolometer and also
stabilizes it so that it stays in its transition.
\paragraph{TES bolometer noise spectra.} 
In Section 3.2.2 we considered the sensitivity of an instrument in the
case of photon occupancy number $n \ll 1$ and mentioned that the
sensitivity is then proportional to $\sqrt{T_{sys}}$, or, more
appropriately because $n \ll 1$ contraindicates the Rayleigh-Jeans
approximation, to $\sqrt{P}$, where $P$ is the power in the incident
radiation field, also known as the photon power. In the general case
for CMB experiments, $n \approx 1$, and so a more general formula is
required (see Reference 33 for a discussion). 

For bolometers, the receiver sensitivity is usually described in terms
of the noiseequivalent power (NEP) that can be measured in a
postdetection bandwidth of 1 Hz. NEP properly has units not of power,
but of $W/\sqrt{Hz}$. A bandwidth of 1 Hz is equivalent to a
half-second of integration time. Besides this factor, converting to
units of $\mu$K$\sqrt{s}$ appropriate for detecting CMB fluctuations
requires the appropriate derivative to convert power to thermodynamic
temperature and requires referencing the NEP to the entrance of the
optics. (In practice, the second point means correcting for the
efficiency $\eta < 1$ in a high-frequency system).  

The total NEP$^2$ for a receiver can be found by summing up the
squares of NEPs from different terms, of which the contribution from
photons, NEP$_\gamma$ , is only one. When NEP$_\gamma$ dominates the
sum, the detector is said to be background limited. Intrinsic sources
of noise in bolometer systems include thermal noise from the photons
transporting heat to the cold bath ($NEP_{th} \approx
\sqrt{2kT^2_0G}$), Johnson noise power in the bolometer resistance
(which is reduced significantly by the electrothermal feedback, as
described in Reference 34), back-end noise (from the SQUIDs), and
occasional other unexplained sources of noise, including $1/f$
noise. 

The inductor in series with the TES resistor serves not only to couple
the TES current to the SQUID output, but also to provide a Nyquist
filter with time constant $L/R_0$. The $L/R_0$ time constant low pass
filters the signals, and the noise power, before they emerge from the
cryostat. Now that we have reviewed the experimental basics and
detectors used for CMB measurements, we turn to the discussion of how
to choose an observing strategy for an experiment.
\subsection{Observing Strategies} 
We are surrounded by a bath of CMB photons, and with all
directions being equal it seems that only astrophysical foregrounds
determine the choice for which parts of the sky should be
observed. The size of the observed patch, together with the angular
resolution, determines the accessible angular scales. However, the
size, shape, and uneven coverage of the observed region, and even the
way in which it is scanned, all impact the determination of the CMB
power spectrum so that an optimization of the observing strategy
requires more sophisticated considerations. A pedagogical illustration
of the choices and their impacts is given in Reference 35. We mention
here the most important issues that must be taken into account in
developing an observing strategy. 

Including the cosmic variance (see Equation 2), the achievable
precision in the power spectrum ($C_l$) can be expressed in the
following form:
\begin{equation}\frac{\Delta C_l}{C_l} = \sqrt{\frac{2}{2l + 1}}\left(\frac{1}{\sqrt{f_{sky}}} + 
\frac{4\pi(T_{exp})^2}{C_l}\sqrt{f_{sky}}e^{l^2 \sigma_b^2}\right).\end{equation} Here,
$f_{sky}$ represents the observed fraction of the sky, $T_{exp}$ the
total experimental sensitivity ($\Delta T$ combining all detectors for
the duration of the run), and $\sigma_b$ the width of the beam. Finite
beam resolution degrades the sensitivity progressively more at higher
$l$ values. For example, the WMAP beam of approximately 0.2$^\circ$
limits sensitivity to below $l \approx 600$; smaller beams are needed
to study much finer scales. In general, the impact of the scanning
strategy on the measurement of the different $C_l$ is summarized in a
window function that describes the weight with which each $C_l$
contributes to the measured temperatures. A crude estimate of the
lowest accessible $l$ is $l_{min} \approx 1/\Delta \Theta$ , with
$\Theta$ being the angular extent of the survey. Limited sky coverage
leads to correlations in the power at different multipoles
$l$. Therefore, the power spectrum is usually reported in largely
uncorrelated $l$ bands of width $\Delta l$, which should be larger
than $l_{min}$. The error on the average $C_l$ in an $l$ band then
follows from Equation 9, with a prefactor of $1/\sqrt{\Delta l}$. 

For an experiment to optimally take advantage of its sensitivity
$\Delta T$, the size of the observed sky patch should be chosen such
that the contributions to the uncertainty in $C_l$ from sample
variance (first summand) and from noise (second summand) are roughly
equal. This leads to $4\pi f_{sky} = C_l e ^{-l^2\sigma_b^2} (\Delta
T)^{-2}$. Another way to express this is that the ideal patch size is
such that the signal-to-noise ratio per pixel $\sigma_{CMB}/(\Delta
T_{pix})$ is unity, where $\sigma_{CMB}$ is the expected CMB
fluctuation in a beam-sized pixel and $T_{pix}$ the experimental error
on the temperature in that pixel. 

A polarization analysis always suffers from small survey sizes because
E and B modes are not local fields, and thus a small patch contains
ambiguous modes. The power spectrum estimation can then lead to E-to-B
leakage and distort the B-mode measurements. Because the CMB B modes
are an order of magnitude smaller than the E modes, even a small
leakage can significantly affect the measurements. However, Smith (36)
has already demonstrated a method for avoiding significant leakage. 

In an optimal experiment, the observed patch is covered homogeneously,
which maximizes the sensitivity to the power spectrum for a given
integration time. It is beneficial to cover any pixel in the observed
patch with different receivers, and also with different receiver
orientations, especially but not only for the polarization
analysis. This cross linking provides robustness against effects from
time-stream filters in the data processing (see Section 3.4.1) and
enables various systematic studies where instrumental effects or
pickup of signals from the ground can be distinguished from real
signals on the sky (see Reference 37 for another view on this). 

Using a purely azimuthal scan on a patch, the atmospheric contribution
is constant, but polarization-sensitive receivers detect different
combinations of $Q$ and $U$ at different times, leading to pixels
having nonuniform weights for $Q$ and $U$. This also compromises the
disentanglement of E and B modes, reducing sensitivity. A more
symmetric distribution is achieved by scanning, for example, in a ring
pattern, resulting in much more uniform $Q/U$ sensitivity and enabling
many systematic cross checks with a high degree of symmetry. However,
a changing atmospheric contribution along the ring must then be taken
into account. 

For any experiment, the scan speed should be as fast as possible to
decrease the effects of $1/f$ drifts (see Section 3.2.2), but upper
limits are posed by the beam resolution, maximum sampling speed,
mechanical constraints of the telescope, and detector time
constants. Typical scanning speeds are tens of arcminutes per
second. Typical sampling at 10-100 Hz then provides several samples
per beam. Given all the above considerations, it is clear that an
optimization of the observing strategy is not trivial and any scan
strategy will include some compromises. So far we have discussed the
challenges in acquiring useful CMB data. In the next section we
introduce the obstacles still lying ahead in the processing of data to
enable access to its cosmological treasures.
\subsection{Techniques of Data Analysis}
In CMB experiments, data are accumulated continuously over a duration
ranging from a few hours (balloon experiments) up to a few years
(space missions). The data volume for past and current experiments was
at most a few hundred gigabytes. Even with planned expansions to
arrays of $\sim 1000$ receivers, the data streams will amount to a
fraction of typical current high-energy physics experiments:
10-50~Tbytes per year. Although for high-energy physics experiments
the data records can be split into different categories of interest
and the analyses are usually performed on specific selections of the
data with an event-wise treatment, here all of the acquired data
contain the same signal and the signal's extraction becomes possible
only with long integration times, requiring the processing of all data
in the same analysis. Typically the CMB analysis can be divided into
four main steps, which are described in the following
sections. 
\subsubsection{Filtering and cutting.} 
Imaging the CMB requires careful cleaning of the data. Selection of
data without instrumental failures or bad weather (for groundbased
experiments) comprises the first filter on the data. The detector time
streams contain long-term drifts of the detector responsivities and
often of the atmosphere, which must be eliminated. Removal of the mean
of the data over short time stretches (10-100 s) or high-pass
filtering is typically used to reduce the effects of such
drifts. Figure 10 demonstrates the impact of drifts on simulations for
the Planck experiment, where the left map was produced without any
filtering and the right map represents the same data but with a
destriping algorithm applied. The scales of the maps are noticeably
different and so is the visible pattern, which emphasizes the need for
such filtering. 

Other components in the data that must be removed are the effects of
ground pickup from the 300-K Earth and changing atmospheric
contributions during the scan. These signals are fixed to the
reference frame of Earth and so are distinguishable from the celestial
CMB signals. Even though any such data filtering removes some
cosmological information, observing strategies are devised so that the
filtering does not significantly compromise the sensitivity.
\begin{figure}
\includegraphics[width=6in]{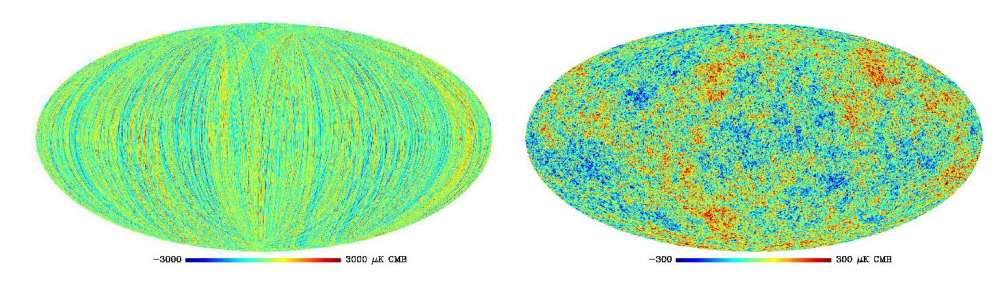}
\caption{Effect of destriping on simulated sky maps. (Left) Map from
a raw time stream. (Right) Map after applying a destriping algorithm
(note the different scales). This simulation was done for the Planck
High Frequency Instrument (38).} 
\end{figure}
\subsubsection{Reduction to maps.} 
The most intuitive representation of the CMB signal is a map on the
sky. If done properly, this provides orders-of-magnitude compression
of the data where no cosmological information is lost (for current
typical experiments, the time streams have $N_t \sim 10^7$ samples
this for 10 or more detectors whereas the number of sky pixels is $N_p
\sim 10^5-10^6$). 

For an ideal time stream containing only the CMB signal on top of
white noise, a map can easily be produced by averaging all
observations that fall into a given pixel on the sky. This makes for a
robust and fast algorithm, which scales linearly with the number of
data samples. However, in a real experiment the filtering of the time
streams introduces correlations that make this simple approach
fail. 

The optimal map estimate from the time stream is produced by
maximizing the likelihood for the data given a certain noise
model. The likelihood problem can be solved analytically, requiring
the inverse of the covariance matrix in the time domain, as well as
the calculation and inversion of the covariance matrix in pixel
space. Whereas the time-stream covariance matrix is sparse, the
covariances in the pixel domain have no special structure and the
matrix inversion dominates the processing time, scaling with
$N_{pix}^3$. Parallelization of this procedure is
possible. Iterative procedures are another approach and can reduce the
required amount of computing power to scale as $N_{it}N_t$~log~$N_t$ ,
where $N_{it}$ is the number of iterations. For those procedures,
Monte Carlo methods are required to estimate the covariance, with a
possible loss in sensitivity.
\subsubsection{Power spectrum estimation.}
Although the power spectrum could in principle be estimated directly
from the time stream, it is more efficient to first produce a map and
from it determine the power spectrum. The likelihood can then be
expressed in the following form: 
\begin{equation} \mathcal{L}= \frac{1}{\sqrt{det(C)}} exp[-{\bf x}^TC^{-1}{\bf x}/2],\end{equation} where
${\bf x}$ represents the map vector and $C$ the pixel-pixel covariance
matrix in which the cosmological information is imbedded. The map
contains contributions from both signal and receiver noise, and,
similarly, the covariance matrix comprises signal added to noise. The
likelihood is a measure of how well the scatter as seen in the data
agrees with that expected from the combination of noise and the CMB
signal. Maximizing the likelihood by setting $\frac{\delta
\mathcal{L}}{C_l} = 0$ leads to an equation that requires an iterative
solution. A common and powerful method for likelihood estimations
exploits a quadratic estimator, which uses for the error matrix an
ensemble average (Fisher matrix) instead of the full curvature (see
Reference 39 for details) and reaches convergence within only a few
iterations. The maximum likelihood method offers an optimal evaluation
of the power spectrum. However, its large matrix operations become
impractical for upcoming experiments with fine-resolution maps of large
fractions of the sky ($>10^6$ pixels). Methods that exploit Monte
Carlo simulations to approximate the analytical solution are thus an
attractive option for the evaluation of the power spectrum and have
become popular [e.g., the pseudo-$C_l$ method (40)]. The use of Monte
Carlo simulations for the evaluation of real data is a familiar
concept to the high-energy physics physicist and is advantageous in
that it enables a relatively easy treatment of various instrumental
and processing artifacts or distortions.
\subsubsection{Cosmological parameter estimation.} 
As described in Section 1, the development and contents of the
Universe determine the characteristics of the CMB anisotropies and
with that the shape of the CMB power spectrum. Current software tools
such as $cmbfast$ or $camb$ calculate the expected power spectrum from
a given set of cosmological parameters within a few seconds so that
for a given measured power spectrum, the likelihood for different
parameter values can be evaluated reasonably quickly. Still, the
likelihood evaluation on a fine grid in the multidimensional parameter
space requires huge computing resources so that the problem is
typically approached via Markov chain Monte Carlo methods. A Markov
chain begins with the evaluation of the likelihood at a specific point
in the parameter space where its value in uences the next point for
the likelihood evaluation. Repeating this leads to a sample density in
parameter space proportional to the likelihood where projections onto
one- or two-dimensional subspaces result in the marginalized
likelihoods. By now, software packages such as $cosmomc$ are available
and in use by the CMB community.
\section{FUTURE PROSPECTS} 
\begin{center}
\begin{table}
\begin{tabular}{llllll}
\hline \hline Signal  & $\Delta T_{cos} (nK) $  & $\Delta T_{Gal} (nK) $ & $\Delta T_{Lens} (nK) $  & $\Delta
 T_{exp}(3\sigma) (nK) $ & $N_{wmap}^{opt}$ \\
\hline lensing, $l=1000$ &  300 & 150   & - & 35  & 1.5 \\
 $r= 0.1$ (SLS) & 83  & 295  & 50  &  9.8  & 20\\
 $r= 0.01$ (SLS) & 26  & 295  & 50  &  3.1 & 200\\
 $r= 0.1$ (reion) & 54  & 780  & - &  6  & 50\\
 $r= 0.01$ (reion) & 17  & 780  & - &  1.9  & 500\\
\hline
\end{tabular}
\caption{Required sensitivities to detect B-mode polarization signals
Signal. For detecting the lensing signal at $l = 1000$ and the
gravity wave at $r = 0.1$ and 0.01, using either the surface of last
scattering or the reionized plasma, we give the magnitude of the
cosmological signal at its peak, the size of the galaxtic
contamination, the magnitude of the lensing contaminant, the derived
total experimental sensitivity to detect the signal (neglecting
foregrounds) at 3$\sigma$, and the corresponding increase over that
achieved with WMAP in 1 year. The foreground estimates are taken from
the WMAP empirical full-sky relation, evaluated at 90 GHz at the
appropriate $l$ value. For small, selected patches of sky, foreground
contamination will be significantly smaller, perhaps by an order of
magnitude. Thus, the lensing signal should be observable from the
ground with just eight WMAPs, for example, an array with eight times
the number of detectors in WMAP, each with the same sensitivity,
observing for one year and so on. The signatures of gravity waves from
the surface of last scattering (SLS) should then also be detectable,
even in the presence of foregrounds, although contamination from
lensing and residual foreground levels will reduce sensitivity from
what is shown in the table. Gravity wave signals from the reionized
plasma have negligible contamination from lensing but, because of the
needed full-sky coverage, larger foregrounds to deal with.}
\end{table}
\end{center}
Here we discuss experiments needed to detect the B-mode signals of
lensing at large $l$ values and of gravity waves at intermediate and
small $l$ values. Table 3 shows the sensitivity required to make
3$\sigma$ detections of several target signals. Estimates come from
Equation 9, either for the full sky or for a smaller patch where the
balance between sample variance and detector noise is optimized.  We
give expressions for the sensitivity to a feature in the power
spectrum centered at $l_0$ and with width $\Delta l = l_0$ and for the
fraction of the sky that accomplishes the balance: ($\Delta
C_l/C_l)_{opt} \approx 3 \Delta T_{exp}/\Delta T_{cos}$ and $f_{opt}
\approx 1/2\left(\frac{\Delta T_{cos}}{l_0 T_{exp}}\right)^2$. For our
purposes, both the signal from lensing and from primordial gravity
waves have approximately this shape. Here $\Delta T_{cos}$ is the
(peak) cosmological signal of interest and $\Delta T_{exp}$ is the
total sensitivity of the experiment, summing over all the observing
time and all the detectors. 

The lensing detection can be accomplished by observing for a year
(from the ground at a good site such as the Atacama Desert in Chile or
the South Pole) a patch of approximately 1.6 square degrees, with
detectors having 1.5 times the WMAP sensitivity. The table also gives
the sensitivities required to detect primordial B modes at various
levels of $r = T/S$. Foregrounds will be a problem for $r \le 0.01$,
perhaps more so for the detection of the signal from the reionized
plasma than from the surface of last scattering. A satellite
experiment can detect the signal from the reionized plasma, where the
lensing contamination is nearly negligible.

The Planck experiment (2007) has an order of magnitude in temperature
sensitivity over WMAP. Polarization sensitivity was not a primary
goal. Still, much has gone into making sure the residual systematic
uncertainties (and foregrounds) can be understood sufficiently well to
allow the extraction of polarization signals around 50 nK,
corresponding to $r = 0.05$. 

There is a program of experiments over the coming five to eight
years. These will involve, progressively, tens, then hundreds, and
nally a thousand or more detectors per experiment and will test
polarization modulation schemes, effective scan strategies,
foreground-removal methods, and algorithms for separating E and B
modes. 

Experiments with tens of detectors are already underway. The sister
experiments QUaD and BICEP observe from the South Pole, using
polarization-sensitive bolometers at 100 and 150 GHz. QUaD, with a
4-arcmin beam, is optimized for gravitational lensing, whereas BICEP,
at approximately 40 arcmin, is searching for gravitational waves. MBI
and its European analog BRAIN are testing the idea of using bolometers
configured as an interferometer, and PAPPA is a balloon effort using
waveguide-coupled bolometers from the Goddard Space Flight
Center. These latter experiments have beams in the range of
0.5$^\circ$ to 1$^\circ$. 

Five initiatives at the level of hundreds of detectors have so far
been put forth. Four use TES bolometers at 90, 150, and 220 GHz:
CLOVER, the lone European effort, with an 8-arcmin beam; Polarbear,
with a 4-arcmin beam; and EBEX and SPIDER, balloon-borne experiments
with 8- and 20 70-arcmin beams, respectively (SPIDER and Polarbear
will use antenna-coupled devices). The fifth uses coherent detectors
at 44 and 90 GHz: QUIET, initially with a 12-arcmin beam, observing
from the Atacama Desert. All are dedicated ground-up polarization
experiments that build their own optical systems. The ACT and SPT
groups, supported by the National Science Foundation, deploy very
large telescopes to study both the cosmology of clusters via the SZ
effect and fine-scale temperature anisotropies, and will likely
propose follow-up polarimeters.
\subsection{The Next Satellite Experiment} 
The reach of the next satellite experiment, termed CMBPOL as defined by
the three-agency task force in the United States (36) and termed BPOL
in Europe, is to detect the signal from gravity waves limited only by
astrophysical foregrounds. Examining Figure 6, we see that $r = 0.01$,
and possibly lower values, can be reached.We should know a great deal
from the suborbital experiments well before the 2018 target launch
date. For studying polarization at large scales, where foregrounds
pose their greatest challenge, information from WMAP and Planck will
be the most valuable.
\section{CONCLUDING REMARKS} 
There is considerable promise for new, important discoveries from the
CMB, ones that can take us back to when the Universe's temperature was
between the Grand Unified Theory and Planck energy scales. This is
particle physics, and while we hope accelerators will provide crucial
evidence for, for example, the particle nature of dark matter,
exploring these scales seems out of their reach. 

In some ways, cosmology has followed the path of particle physics: It
has its Standard Model, accounting for all confirmed phenomena. With no
compelling theory, parameter values are not of crucial interest. We
cannot predict the mass of the top quark, nor can we predict the
primordial energy densities. Each discipline is checking consistency,
as any discrepancies would be a hint of new physics. 

The CMB field is not as mature as particle physics. It needs
considerable detector development, even for current experiments. There
is rapid progress, and overall sensitivity continues to
increase. Foregrounds are certainly not sufficiently known or
characterized. There is a great deal of competition in the CMB, like
the early days of particle physics before the experiments grew so
large that more than one or two teams exploring the same topic
worldwide was too costly. For the moment, this is good, as each team
brings something unique in terms of control of systematics,
frequencies, regions of the sky scanned, and detection
technology. However, there is a difference in the way results are
reported in the two fields. In the CMB field, typically almost nothing
is said about an experiment between when it is funded and when it
publishes. Here publishing means that results are announced, multiple
papers are submitted and circulated, and often there is a full data
release, including not only of raw data and intermediate data products
but sometimes support for others to repeat or extend the analyses. The
positives of this tradition are obvious. However, one negative is that
one does not learn the problems an experiment is facing in a timely
manner. There is a degree of secrecy among CMB scientists. 

There are other differences. CMB teams frequently engage theorists to
perform the final analysis that yields the cosmological significance of
the data.  Sophisticated analysis techniques are being developed by a
set of scientists and their students who do not work with detectors
but do generate a growing literature. There are as yet no standardized
analysis techniques; effectively each new experiment invents its
own. The days appear to be over where the group of scientists that
design, build, and operate an experiment can, by themselves, do the
full scientific analysis. Another distinction is that there is no one
body looking over the field or advising the funding agencies, and
private funds sometimes have a major impact. Nearly all CMB scientists
are working on multiple projects, sometimes as many as four or five,
holding that many grants. More time is spent writing proposals and
reports and arranging support for junior scientists, for whom there is
little funding outside of project funds. This is certainly not the
optimum way to fund such an exciting and promising field.
\section{ACKNOWLEDGMENTS} 
The authors have enjoyed many productive conversations with colleagues
at their institutions in the course of preparing this review. The
authors are particularly grateful for helpful comments from Norman
Jarosik, Bernd Klein, Laura La Porta, and Kendrick Smith. We also wish
to acknowledge our collaborators in QUIET and CAPMAP for frequent
discussions of relevant issues. This work was partially supported by
grants from the National Science Foundation: PHY-0355328, PHY-
0551142, and ASTR/AST-0506648.
\section*{References}
\begin{itemize}
\item[1.] Kamionkowski M, Kosowsky A. Annu. Rev. Nucl. Part. Sci. 49:77 (1999) 
\item[2.] Dodelson S. Modern Cosmology. Acad. Press (2003) 
\item[3.] Penzias AA,Wilson RW. Astrophys. J. 142:419 (1965) 
\item[4.] Partridge RB. 3K: The Cosmic Microwave Background Radiation. Cambridge, NY: Cambridge Univ. Press (1995) 
\item[5.] Fixsen DJ, et al. Astrophys. J. 483:586 (1996) 
\item[6.] Feng JL, Rajaraman A, Takayama F. Phys. Rev. D 68:063504 (2003) 
\item[7.] Smoot GF, et al. Astrophys. J. 396:1 (1992) 
\item[8.] Hinshaw G, et al. Astrophys. J. In press (2007) 
\item[9.] Bock J, et al. astro-ph/0604101 (2006) 
\item[10.] Smith TL, Kamionkowski M, Cooray A. Phys. Rev. D 73(2):023504 (2006) 
\item[11.] Spergel DN, et al. Astrophys. J. In press (2007) 
\item[12.] Page L, et al. Astrophys. J. Suppl. 148:233 (2003) 
\item[13.] Readhead ACS, et al. Science 306:836 (2004) 
\item[14.] MacTavish CJ, et al. Astrophys. J. 647:799 (2006) 
\item[15.] Tegmark M, et al. Phys. Rev. D 69:103501 (2004) 
\item[16.] Hu W, Sugiyama N. Phys. Rev. D 51:2599 (1995) 
\item[17.] Eisenstein DJ, et al. Astrophys. J. 633:560 (2005) 
\item[18.] Bennett CL, et al. Astrophys. J. Suppl. 148:97 (2003) 
\item[19.] Page L, et al. Astrophys. J. In press (2007) 
\item[20.] La Porta L, Burigana C, Reich W, Reich P. Astron. Astrophys. 455(2):L9 (2006) 
\item[21.] Carretti E, Bernardi G, Cortiglioni S. MNRAS 373:93 (2006) 
\item[22.] Hanany S, Rosenkranz P. New Astron. Rev. 47(11 12):1159 (2003) 
\item[23.] Pietranera L, et al. MNRAS. In press (2007) 
\item[24.] Finkbeiner DP, Davis M, Schlegel DJ. Astrophys. J. 524:867 (1999) 
\item[25.] Vaillancourt JE. Eur. Astron. Soc. Publ. Ser. 23:147 (2007) 
\item[26.] Benoit A, et al. Astron. Astrophys. 424:571 (2004) 
\item[27.] Ponthieu N, et al. Astron. Astrophys. 444:327 (2005) 
\item[28.] Huffenberger KM, Eriksen HK, Hansen FK. Astrophys. J. 651:81 (2006) 
\item[29.] Stivoli F, Baccigalupi C, Maino D, Stompor R. MNRAS 372:615 (2006) 
\item[30.] Verde L, Peiris HV, Jimenez R. J. Cosmol. Astropart. Phys. 1:19 (2006) 
\item[31.] Barkats D, et al. Astrophys. J. Suppl. 159(1):1 (2005) 
\item[32.] Ryle M. Proc. R. Soc. A 211:351 (1952) 
\item[33.] Zmuidzinas J. Appl. Optics 42:4989 (2003) 
\item[34.] Mather JC. Appl. Optics 21:1125 (1982) 
\item[35.] Tegmark M. Phys. Rev. D 56:4514 (1997) 
\item[36.] Smith K. New Astron. Rev. 50(11 12):1025 (2006) 
\item[37.] Crawford T. astro-ph/0702608 (2007) 
\item[38.] Couchout F, et al. CMB and physics of the early universe. http://pos.sissa.it/ cgi-bin/reader/conf.cgi?confid=27 (2006) 
\item[39.] Bond JR, Jaffe AH, Knox L. Astrophys. J. 533:19 (2000) 
\item[40.] Hivon E, et al. Astrophys. J. 567:2 (2002)
\end{itemize}

\end{document}